\documentclass[journal]{IEEEtran}
\usepackage[mathscr]{eucal}
\usepackage[cmex10]{amsmath}
\usepackage{epsfig,epsf,psfrag}
\usepackage{amssymb,amsmath,amsthm,amsfonts,latexsym}
\usepackage{amsmath,graphicx,bm,xcolor,url,overpic}
\usepackage{fixltx2e}
\usepackage{array}
\usepackage{verbatim}
\usepackage{bm}
\usepackage{algorithmic}
\usepackage{algorithm}
\usepackage{verbatim}
\usepackage{textcomp}
\usepackage{mathrsfs}
\usepackage{epstopdf}

\newcommand{\openone}{\leavevmode\hbox{\small1\normalsize\kern-.33em1}}

\catcode`~=11 \def\UrlSpecials{\do\~{\kern -.15em\lower .7ex\hbox{~}\kern .04em}} \catcode`~=13 

\allowdisplaybreaks[4]

\newcommand{\nn}{\nonumber}

\newcommand{\calC}{\mathcal{C}}
\newcommand{\calD}{\mathcal{D}}
\newcommand{\calE}{\mathcal{E}}

\newcommand{\calL}{\mathcal{L}}
\newcommand{\calM}{\mathcal{M}}

\newcommand{\calQ}{\mathcal{Q}}

\newcommand{\calT}{\mathcal{T}}
\newcommand{\calU}{\mathcal{U}}

\newcommand{\calX}{\mathcal{X}}
\newcommand{\calY}{\mathcal{Y}}
\newcommand{\calZ}{\mathcal{Z}}

\newcommand{\rma}{\mathrm{a}}

\newcommand{\rmc}{\mathrm{c}}

\newcommand{\rmt}{\mathrm{t}}

\newcommand{\rmu}{\mathrm{u}}

\DeclareMathAlphabet{\mathbsf}{OT1}{cmss}{bx}{n}
\DeclareMathAlphabet{\mathssf}{OT1}{cmss}{m}{sl}

\DeclareSymbolFont{bsfletters}{OT1}{cmss}{bx}{n}  
\DeclareSymbolFont{ssfletters}{OT1}{cmss}{m}{n}
\DeclareMathSymbol{\bsfGamma}{0}{bsfletters}{'000}
\DeclareMathSymbol{\ssfGamma}{0}{ssfletters}{'000}
\DeclareMathSymbol{\bsfDelta}{0}{bsfletters}{'001}
\DeclareMathSymbol{\ssfDelta}{0}{ssfletters}{'001}
\DeclareMathSymbol{\bsfTheta}{0}{bsfletters}{'002}
\DeclareMathSymbol{\ssfTheta}{0}{ssfletters}{'002}
\DeclareMathSymbol{\bsfLambda}{0}{bsfletters}{'003}
\DeclareMathSymbol{\ssfLambda}{0}{ssfletters}{'003}
\DeclareMathSymbol{\bsfXi}{0}{bsfletters}{'004}
\DeclareMathSymbol{\ssfXi}{0}{ssfletters}{'004}
\DeclareMathSymbol{\bsfPi}{0}{bsfletters}{'005}
\DeclareMathSymbol{\ssfPi}{0}{ssfletters}{'005}
\DeclareMathSymbol{\bsfSigma}{0}{bsfletters}{'006}
\DeclareMathSymbol{\ssfSigma}{0}{ssfletters}{'006}
\DeclareMathSymbol{\bsfUpsilon}{0}{bsfletters}{'007}
\DeclareMathSymbol{\ssfUpsilon}{0}{ssfletters}{'007}
\DeclareMathSymbol{\bsfPhi}{0}{bsfletters}{'010}
\DeclareMathSymbol{\ssfPhi}{0}{ssfletters}{'010}
\DeclareMathSymbol{\bsfPsi}{0}{bsfletters}{'011}
\DeclareMathSymbol{\ssfPsi}{0}{ssfletters}{'011}
\DeclareMathSymbol{\bsfOmega}{0}{bsfletters}{'012}
\DeclareMathSymbol{\ssfOmega}{0}{ssfletters}{'012}

\newcommand{\tilE}{\tilde{E}}

\newcommand{\tilm}{\tilde{m}}

\newcommand{\hatQ}{\hat{Q}}

\newcommand{\tilQ}{\tilde{Q}}

\newcommand{\tilu}{\tilde{u}}

\newcommand{\barQ}{\bar{Q}}

\newcommand{\iid}{i.i.d.\ }

\newcommand{\dotleq}{\stackrel{.}{\leq}}

\newcommand{\markov}{\mathrel{\multimap}\joinrel\mathrel{-}%
\joinrel\mathrel{\mkern-6mu}\joinrel\mathrel{-}}

\newtheorem{theorem}{Theorem} 
\newtheorem{lemma}[theorem]{Lemma}

\newtheorem{proposition}[theorem]{Proposition}

\newtheorem{remark}{Remark}
\newtheorem{fact}{Fact}

\usepackage{cite,xcolor}
\usepackage[ colorlinks = true,
       linkcolor = blue,
       urlcolor = blue,
       citecolor = red,
       anchorcolor = green,]{hyperref}
\usepackage{amsmath,amssymb,amsfonts,url}
\usepackage{algorithmic}
\usepackage{enumitem}
\usepackage{graphicx}
\usepackage{textcomp}
\usepackage{bbm,overpic}
\def\BibTeX{{\rm B\kern-.05em{\sc i\kern-.025em b}\kern-.08em
    T\kern-.1667em\lower.7ex\hbox{E}\kern-.125emX}}

\begin{document}
\flushbottom
\allowdisplaybreaks[1]

\title{Exact Error and Erasure Exponents for the Asymmetric Broadcast Channel}
\author{
  \IEEEauthorblockN{Daming Cao, {\em Student Member, IEEE}} $\qquad$ 
   \IEEEauthorblockN{Vincent Y.~F.~Tan, {\em Senior Member, IEEE}}
 \thanks{D.\ Cao is with the Southeast University of China
 (e-mail: dmcao@seu.edu.cn). }
   \thanks{V.\ Y.\ F.\ Tan is with  the National University of Singapore
 (e-mail: vtan@nus.edu.sg). } \thanks{D.\ Cao is  supported by the China Scholarship  Council (No.\ 201706090064) and the National Natural Science Foundation of China under Grant No.\ 61571122. V.~Y.~F.~Tan is supported by a Singapore National Research Foundation (NRF) Fellowship (R-263-000-D02-281).} \thanks{This paper was presented in part at the 2018 IEEE International Symposium on Information Theory~\cite{CaoTan18}.  } \thanks{Copyright (c) 2017 IEEE. Personal use of this material is permitted.  However, permission to use this material for any other purposes must be obtained from the IEEE by sending a request to pubs-permissions@ieee.org.}
} 
\maketitle

\begin{abstract}
Consider the asymmetric broadcast channel with a random superposition codebook, which may be comprised of constant composition or \iid codewords. By applying Forney's optimal decoder for individual messages and the message pair for the receiver that decodes both messages,  exact (ensemble-tight) error and erasure exponents are derived.  It is shown that the optimal decoder designed to decode the pair of messages   achieves the optimal trade-off between the  total and undetected   exponents associated with the optimal decoder for the private message.  Convex optimization-based procedures to evaluate the exponents efficiently are proposed. Finally, numerical examples are presented to illustrate the results. 
\end{abstract}

\begin{IEEEkeywords}
Broadcast channels, Degraded Message Sets, Erasure decoding, Undetected Error, Error exponents, Superposition coding.
\end{IEEEkeywords}
\section{Introduction} 

\subsection{Background and Related Works}
The broadcast channel~\cite{cover1972broadcast}    has been extensively studied   in multi-user information theory. Although the capacity region is still unknown, some special cases have been solved. One   example is the broadcast channel with degraded message sets, also known as the asymmetric broadcast channel (ABC). For this channel, one receiver desires to decode both the private message  $m_1$ and the common message $m_2$ while the other  receiver desires to decode  only~$m_2$.  This model can be applied to a plethora of different scenarios; see Section~\ref{sec:ex} for   concrete examples of   broadcasting scenarios, taking into account the variation we consider herein. 

The capacity region for the  ABC was derived by K\"orner and Marton and is well known~\cite{korner1977general}.   The earliest work on error exponents for the ABC  is that by  K\"orner and Sgarro~\cite{korner1980universally}, who used a constant composition ensemble for deriving an achievable error exponent. Later, Kaspi and Merhav~\cite{kaspi2011error} improved this work by deriving a tighter lower bound for the error exponent by analyzing the ensemble of \iid random codes. Most recently, Averbuch {\em et al.}  derived the exact random coding error exponents and expurgated exponents for the ensemble of constant composition codes in~\cite{averbuch2018exact} and~\cite{averbuch2017expurgated}, respectively.

In this paper, we are interested in decoders with an erasure option. In this setting,  the decoders may, instead of declaring that a particular  message or set of messages is sent,  output an erasure symbol. For the discrete memoryless channel (DMC), Forney~\cite{forney1968exponential}   found the optimal decoder  and derived  a lower bound on  the total and undetected error exponents using Gallager-style bounding techniques. Csisz\'ar and K\"orner~\cite[Thm.~10.11]{csiszar2011information} derived universally attainable erasure and error exponents using a generalization of the maximum mutual information (MMI) decoder. Telatar~\cite{telatar1992multi} also analyzed an erasure decoding rule with a general decoding metric. Moulin~\cite{moulin2009neyman} generalized this family of  decoders and proposed a new decoder parameterized by a weighting function.  Merhav~\cite{merhav2008error} derived lower bounds to these exponents by using a novel type-class enumerator method. In a breakthrough, Somekh-Baruch and Merhav~\cite{somekh2011exact} derived the {\em exact} random coding exponents for erasure decoding. Recently, Huleihel {\em et al.}~\cite{huleihel2016erasure} showed that the  random coding exponent for erasure decoding is not universally achievable and established a simple relation between the total and undetected error exponents. Weinberger and Merhav~\cite{weinberger2017simplified} analyzed a simplified decoder for erasure decoding. Hayashi and Tan~\cite{Hayashi2015asymmetric} derived ensemble-tight moderate deviations and second-order results for erasure decoding over additive DMCs. For the ABC, Tan~\cite{tan2015error} derived lower bounds on the total and undetected error exponents of an extended version of the universal decoder in Csisz\'ar and K\"orner~\cite[Thm.~10.11]{csiszar2011information}. Moreover, Merhav in another landmark work in~\cite{merhav2014exact} analyzed a random coding scheme with a binning (superposition coding)  structure and  showed that a potentially suboptimal bin index decoder achieves the random coding error exponent for decoding only the bin index.

\begin{figure*}[t]
\centering
\setlength{\unitlength}{0.06cm}
\begin{picture}(200,120)
\thicklines
\put(0,10){\vector(1,0){200}}
\put(10,00){\vector(0,1){120}}
\put(185,2){Threshold $T$}
\put(0,90){\rotatebox{90}{Exponents}} 
\color{blue}{
\qbezier(10,60)(85,10)(160,10)
\qbezier(10,60)(85,60)(160,100)
\put(163,13){$E_1^{\rmu}=E_Y^{\rmu}$}
\put(163,100){$E_1^{\rmt}=E_Y^{\rmt}$}
}
\color{red}{
\qbezier(10,80)(85,30)(160,30)
\qbezier(10,80)(85,80)(160,120) 
\put(163,120){$E_2^{\rmu}$}
\put(163,30){$E_2^{\rmt}$}
}
\color{black}{
\put(100,17){\vector(0,1){58}} 
\put(100,75){\vector(0,-1){58}}
\put(103,45){$T$} 

\put(120,33){\vector(0,1){69}} 
\put(120,102){\vector(0,-1){69}} 
\put(123,67){$T$} 
}
\end{picture}
\caption{For a fixed rate pair $(R_1, R_2)$ with varying threshold $T$, the figure schematically illustrates  $E_1^{\rmu}, E_1^{\rmt}, E_2^{\rmu}, E_2^{\rmt}, E_Y^{\rmu}$, and $ E_Y^{\rmt}$ which are respectively, the undetected exponent for decoding $m_1$, the total exponent for decoding $m_1$, the undetected exponent for decoding $m_2$, the total exponent for decoding $m_2$, the undetected exponent for decoding $(m_1, m_2)$,  and the total exponent for decoding $(m_1,m_2)$.   Note that $E_1^{\rmu}=E_Y^{\rmu}$  and $E_1^{\rmt}=E_Y^{\rmt}$ so decoding $m_1$ optimally and $(m_1, m_2)$ optimally result in the same undetected-total exponent trade-off. Clearly, the same is not true of optimal decoding of $m_2$ and optimal decoding of  $(m_1, m_2)$.}
\label{fig:scheme}
\end{figure*}
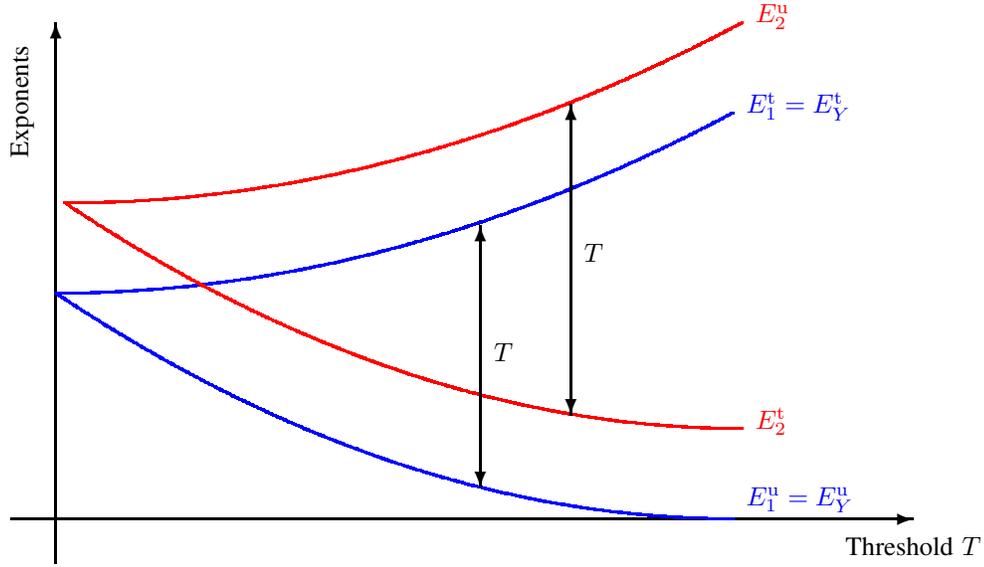
\subsection{Main Contributions} \label{sec:main_contributions}
In this paper, we consider  erasure decoding for the ABC with a  superposition codebook structure, in which the distribution of the codewords is either \iid or constant composition.  For the decoder that aims to decode both messages, there are six exponents of interest---the total and undetected exponents corresponding to the individual messages $m_1$ and $m_2$ and the pair of messages $(m_1,m_2)$. We derive exact (ensemble-tight) exponents for this problem. The main technical contribution to obtain the exact random coding exponents is a set of tools to handle statistical dependencies between codewords that share the same cloud center. To wit, Lemmas~\ref{lem:dependency} and \ref{lem:remove_dependency} consists of two technical results that are to establish the equality between the total random coding error exponents pertaining to the first  message (i.e., the private message $m_1$) and the message pair.  This ameliorates the dependency problem, at least on the exponential scale, which is the asymptotic regime of interest.

We show that the minimizations required to evaluate these error exponents can be cast  as convex optimization problems, and thus, can be solved efficiently using off-the-shelf convex optimization solvers such as CVX. As such, it is computationally tractable to compare the performance of practical codes to the information-theoretic limits presented here; this guides the design and analysis of future generations of   codes. We   present  numerical examples to illustrate these exponents and the trade-offs involved in the erasure decoding problem for the ABC. We  additionally show that the constant composition exponents are, in general, larger than the \iid exponents.

\subsection{Motivation, Significance,   Insights Gleaned, and a Surprise} \label{sec:mot}

Our {\em motivation}  is to find   exact (ensemble-tight) erasure and error exponents for the ABC and from the resulting form of the exponents,   hope to gain valuable insights into the various trade-offs that are present. In particular, we are interested in whether the optimal decoder for the pair of messages $(m_1, m_2)$ (at the receiver that is required to decode both messages) performs as well as that for decoding only the private message  $m_1$ or, for that matter, the common message $m_2$. Our main  observation is that the optimal decoder for $(m_1, m_2)$ achieves the optimal trade-off between the total and undetected exponents pertaining to   $m_1$.   What are the practical engineering implications and significance of this finding? In a   broadcasting   setting, the punchline of this paper says  that if a communication engineer has the erasure option---e.g., in automatic repeat request/query (ARQ)~\cite{forney1968exponential} systems---and desires to only to decipher the private message $m_1$, she can essentially obtain the other (common) message $m_2$ {\em for free} using a decoder designed to decode {\em both} $m_1$ {\em and}  $m_2$. By ``for free'', we mean that the optimal trade-off in the total and undetected exponents  for---i.e., the {\em performance} of---decoding $m_1$ is the {\em same} as that for $(m_1, m_2)$. In view of the {\em packing lemma} \cite[Lemma~3.1]{el2011network} as applied to broadcast channels~\cite[Chapters 5 and 8]{el2011network}, this observation might seem natural or unsurprising in the rate or capacity sense. However, what we show is much more---indeed, a refined asymptotic result. Our main observation and {\em insight gleaned}, which {\em is surprising}, implies that {\em on the exponential scale}---i.e., in terms of error and erasure exponents---there is no loss in the trade-off whether we choose to decode $m_1$ or $(m_1, m_2)$. On the other hand, if the engineer desires to  decode {\em only} $m_2$, she needs to design a {\em dedicated} decoder for this task since the optimal trade-off in the total and undetected exponents for the joint decoder is, in general, worse than that of the dedicated one for $m_2$. This is illustrated schematically in Figure~\ref{fig:scheme}.

\subsection{Practical, Real-Life Examples } \label{sec:ex}

Let us provide practical, real-life {\em examples} for which the above theoretical observation is applicable. 

\begin{figure*}[t]
  \centering
   \begin{overpic}[width=.475\textwidth]{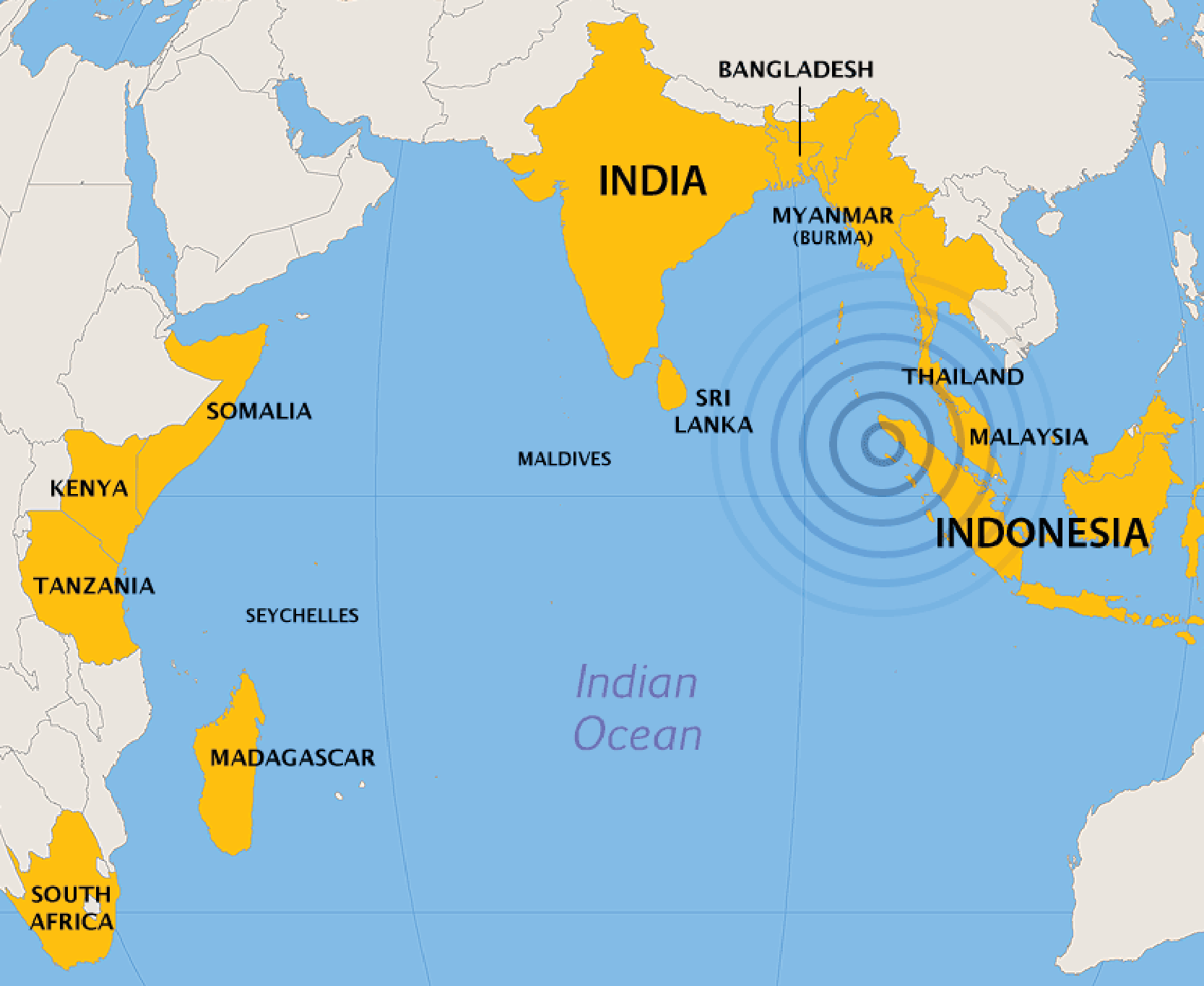}
   \thicklines
    \put(-10,23){\vector(1,1){30}}
    \put(-42,18){\mbox{Needs to decode {\em only}}}
    \put(-42,13){\mbox{location of epicenter $m_2$}}
   \put(115,5){\vector(-1,1){40}}
       \put(116,12){\mbox{Needs to decode {\em both}}}
    \put(116,7){\mbox{location and actions}}
    \put(116,2){\mbox{$(m_1, m_2)$}}
   \end{overpic}
     \caption{A map of countries affected by the 2004 Indian Ocean tsunami and its epicenter. In Sumatra, both $m_1 $ and $m_2 $ should be broadcasted to the residents. Decoding both is just as reliable, in the sense of total and undetected exponents, as decoding only $m_1$. Thus, the main take-home message  of this paper is that $m_2$ {\em comes for free} (i.e., $E_1^{\rmt}=E_Y^{\rmt}$ and $E_1^{\rmu}=E_Y^{\rmu}$ in Figure~\ref{fig:scheme}). In Somalia, though, $m_2$, the location of the tsunami's epicenter is the more salient piece of information and $m_1$ does not have to be known to the populace since they are unlikely to be required to take any substantial action. The performance of decoding $m_2$ alone is not the same as decoding $(m_1, m_2)$ and a dedicated decoder should be designed for the former (i.e., for Somalia). The figure, apart from annotations, is taken from \url{https://en.wikipedia.org/wiki/2004_Indian_Ocean_earthquake_and_tsunami}.}
     \label{fig:tsu}
\end{figure*}
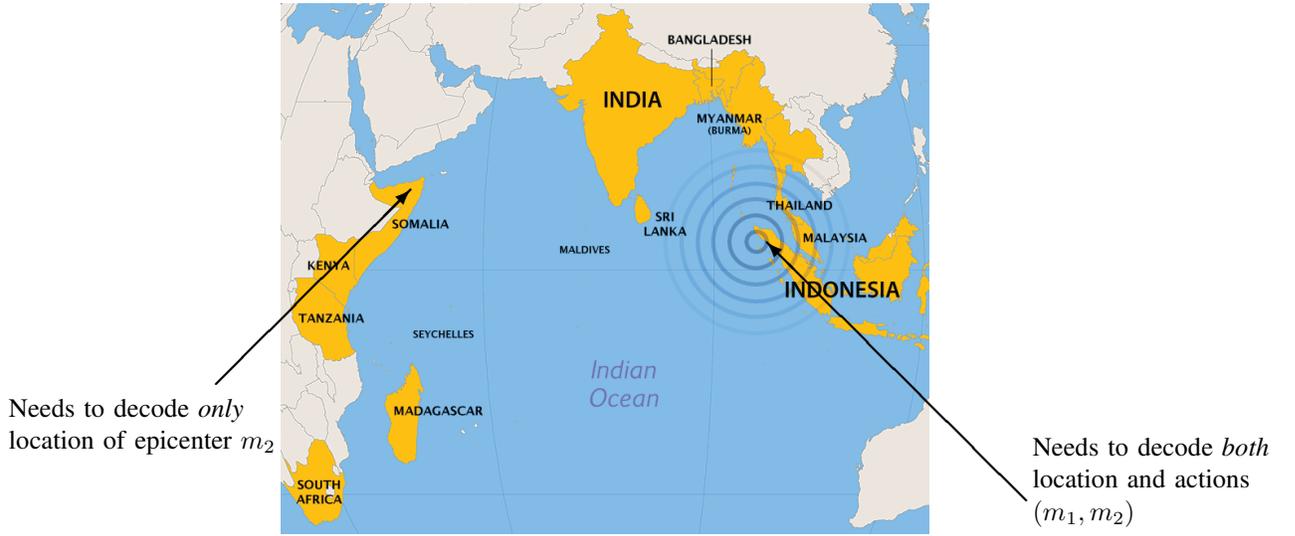

On Boxing Day in 2004, the massive Indian Ocean earthquake and tsunami struck. Its epicenter was off the west coast of northern Sumatra, Indonesia. This event resulted in a tremendous loss of lives (roughly a quarter million) and property (roughly worth USD \$15 billion) to Indonesia, Sri Lanka, Myanmar, Thailand, the Maldives, and even countries as far as Somalia in East Africa in which damage was present but markedly less severe. See Figure~\ref{fig:tsu}. Since then, tsunami warning systems have been set up in Indonesia  among other countries. These warning systems (such as DART\,\textsuperscript{\footnotesize\textregistered}~or \underline{D}eep-ocean \underline{A}ssessment and
\underline{R}eporting of \underline{T}sunamis) are used  detect tsunamis in advance and to issue warnings to people that might be adversely affected; see \cite{Bernard11,Bernard15}. Often, various disparate pieces of information need to be disseminated or broadcast to common folk reliably. For example, those in the direct path of the tsunami may need to know $m_1$, the {\em actions} they should take to avoid loss of lives  (e.g., move to higher ground) and $m_2$,  the  {\em locations} in which the tsunami will make landfall and the corresponding {\em severities}. For such countries, our result says that if the optimal decoder for $(m_1,m_2)$ is used, the performance, as defined in Section~\ref{sec:mot}, is the same as that for decoding only $m_1$. Hence,   the take-home message is that the residents of Sumatra will, in addition to the actions they need to take, also know the   locations the  tsunami makes landfall. This can be done  {\em without any loss of optimality from the perspective of the trade-off between the error and erasure exponents}. For countries that are far away from a major fault line such as Somalia, perhaps the design of a decoder for only $m_2$ is needed since the presence of the tsunami in Southeast Asia is not likely to require any drastic action from   Somalians, so information about $m_1$ is not needed there. In this case, the Somalian authorities and engineers need to design a {\em dedicated} decoder to ensure optimality of decoding $m_2$ with the erasure option. Note that since tsunami warning systems have the potential to save hundreds of thousands of lives, they have to be ultra reliable. As such, our error and erasure formulation, in which the undetected exponents are designed to be   larger than their erasure counterparts (and hence the undetected error probability is exponentially smaller than its erasure counterpart), is of particular relevance in this critical setting. In sum, the findings of our paper have the potential to guide the design and analysis of ultra-reliable   infrastructure with varying demands, such as next-generation tsunami warning systems.

\begin{figure}[t]
  \centering
   \begin{overpic}[width=.45\textwidth]{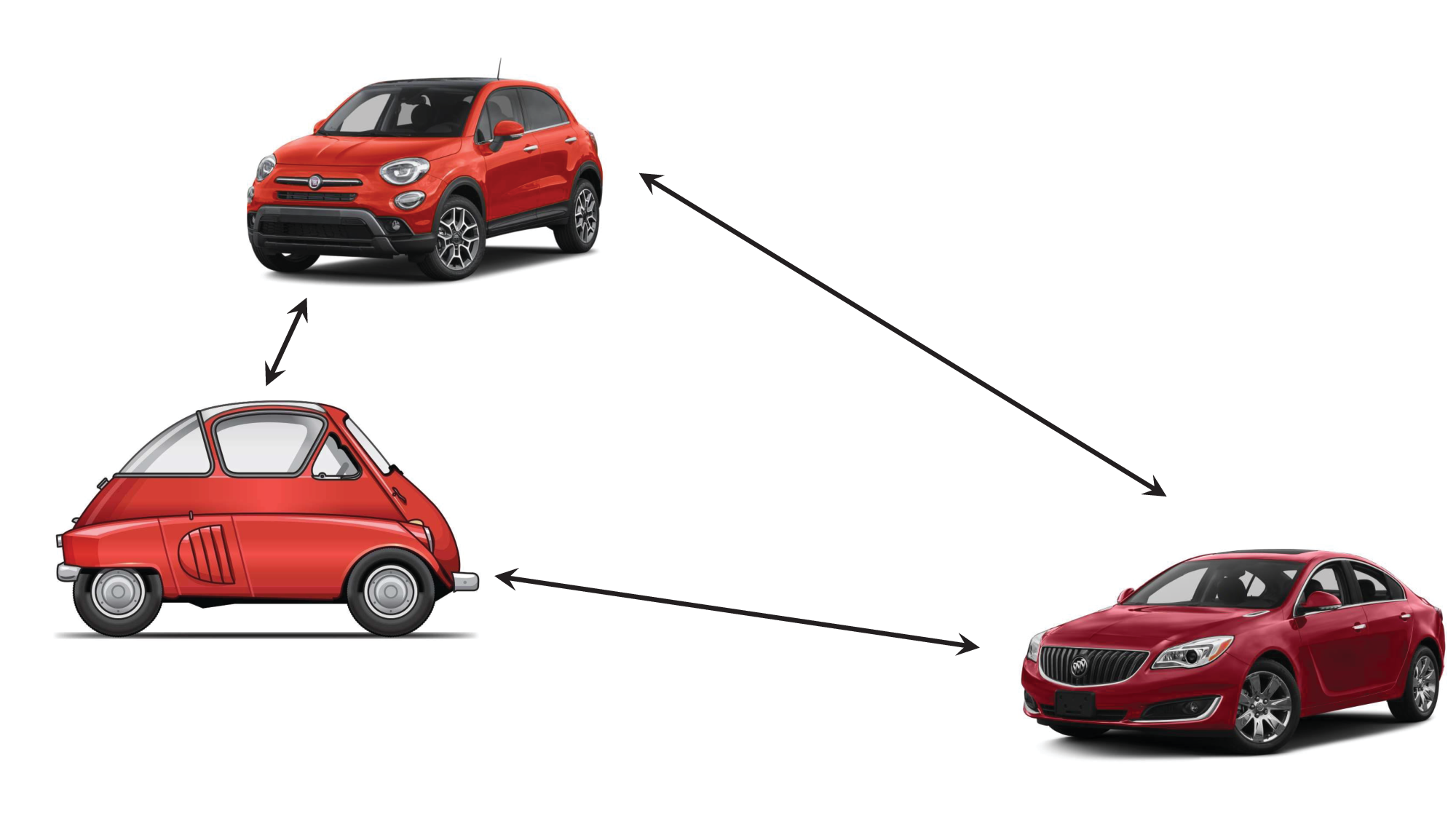}
   \thicklines
    \put(-3,20){\mbox{$\calX$}}
\put(42,50){\mbox{$\calY$ needs $(m_1, m_2)$}}
\put(82,22){\mbox{$\calZ$ needs $m_2$}}
   \end{overpic}
     \caption{A V2V communication network~\cite{arena19} in which receivers have different demands }
     \label{fig:cars}
\end{figure}

Another example comes from vehicle-to-vehicle (V2V) communications~\cite{arena19}. In these systems, vehicles form a communication network in which the vehicles themselves are communicating nodes. Through wireless transmissions, they provide each other with crucial information to enhance the safety of all vehicles involved and, in particular, to prevent accidents. For a concrete example, let us consider three vehicles $\calX$, $\calY$ and $\calZ$; see Figure~\ref{fig:cars}. $\calX$ is in close proximity to $\calY$  and thus the two vehicles are likely to collide if no further action is taken. On the other hand, $\calZ$ is farther away from $\calX$ than $\calY$ is. Hence, in this ultra-reliable setting, $\calX$ desires to transmit $m_1$, the course of actions $\calY$ should take to avoid the crash and $m_2$, its own location. The {\em good news} from our result says that  using the optimal decoder, there is no loss in optimality in decoding both messages vis-\`a-vis only   $m_1$. Since $\calZ$ does not need to take any actions at this point in time, it does not need to know $m_1$ and instead only needs to decode $m_2$.  The optimal decoder for $\calZ$ needs to be designed differently from that for $m_1$ and $(m_1, m_2)$.

%



\section{Problem Formulation}
\subsection{Notation}
Throughout this paper, random variables (RVs) will be denoted by upper case letters, their specific values will be denoted by the respective lower case letters, and their alphabets will be denoted by calligraphic letters. A similar convention will apply to random vectors of dimension $n \in\mathbb{N}$ and their realizations.  For example, the random vector $X^n=(X_1,\dots,X_n)$ may take on a certain realization  $x^n=(x_1,\dots,x_n)$ in $\mathcal{X}^n$, the $n$-th order Cartesian power of $\mathcal{X}$, which is the alphabet of each component of this vector.

The distributions associated with random variables will be denoted by the letters $P$ or $Q$, with subscripts being the names of the random variables, e.g., $Q_{UXY}$ stands for a joint distribution of a triple of random variables $(U,X,Y)$ on $\mathcal{U}\times\mathcal{X}\times\mathcal{Y}$, the Cartesian product alphabets of $\mathcal{U}$, $\mathcal{X}$ and $\mathcal{Y}$. In accordance with these notations, the joint distribution induced by $Q_Y$ and $Q_{X|Y}$ will be denoted by $Q_{XY}:= Q_{Y}Q_{X|Y}$. Information measures induced by the joint distribution $Q_{XY}$ (or $Q$ for short) will be subscripted by $Q$. For example, $I_{Q}(X;Y)$ denotes the mutual information of the random variables $X$ and $Y$ with joint distribution $Q=Q_{XY}$.

For a sequence $x^n$, let $\hat{P}_{x^n}$ denote its empirical distribution or type. The type class $\mathcal{T}_{P_X}$ of $P_X$ is the set of all $x^n$ whose empirical distribution is $P_X$. For a given conditional probability distribution $P_{X|U}$ and sequence $u^n$, $\mathcal{T}_{P_{X|U}}(u^n)$ denotes the conditional type class of $x^n$ ($P_{X|U}$-shell) given $u^n$, namely, the set of sequences $x^n$ whose joint empirical distribution with $u^n$ is given by $P_{X|U}\hat{P}_{u^n}$.

The probability of an event $\mathcal{E}$ will be denoted by $\Pr\{\mathcal{E}\}$, and the expectation operator with respect to a joint distribution $Q$, will be denoted by $\mathbb{E}_{Q}\{\cdot\}$. For two positive sequences $\{a_n\}$ and $\{b_n\}$, the notation $a_n\doteq b_n$ means that $\{a_n\}$ and $\{b_n\}$ are of the same exponential order, i.e., $\lim_{n\to \infty}\frac{1}{n}\ln\frac{a_n}{b_n}=0$. 
Similarly, $a_n\stackrel{\cdot}{\le} b_n$ means that $\limsup_{n\to \infty} \frac{1}{n}\ln\frac{a_n}{b_n}\leq 0$. The indicator function of an event $\mathcal{E}$ will be denoted by $\mathbbm{1}\{\mathcal{E}\}$. The notation $|x|_{+}$ will stand for $\max\{x,0\}$ and notation $[M]$ stands for $\{1,\ldots,M\}$. Finally, logarithms and exponents will be understood to be taken to the natural base. 
\subsection{System Model} \label{sec:model}
We consider a discrete memoryless ABC $\mathcal{W}:\mathcal{X}\to \mathcal{Y}\times\mathcal{Z}$ with a finite input alphabet $\mathcal{X}$, finite output alphabets $\mathcal{Y}$ and $\mathcal{Z}$ and a transition probability matrix $\{W(y,z|x) : x\in \mathcal{X},y\in \mathcal{Y},z\in \mathcal{Z}\}$. Let $W_\mathcal{Y}: \mathcal{X}\to \mathcal{Y}$ and $W_\mathcal{Z}: \mathcal{X}\to\mathcal{Z}$ be respectively the $\mathcal{Y}$- and $\mathcal{Z}$-marginals of $W$.\par 
Assume there is a random codebook $\mathcal{C}$ with superposition structure for this ABC, where the message pair $(m_1,m_2)$ is destined  for user $\mathcal{Y}$ and the common message $m_2$ is destined for user $\mathcal{Z}$. In this paper, we consider i.i.d. random codes and constant composition random codes.
\begin{itemize}
\item For \iid random codes, fix a distribution $P_{UX}(u,x)$ and randomly generate $M_2=e^{nR_2}$ ``cloud centers'' $\{U^n(m_2) : m_2\in \mathcal{M}_2=[M_2]\}$ according to the distribution 
\begin{equation}\label{def:Uiid}
P(u^n):=\prod_{i=1}^n P_U(u_i).
\end{equation}
For each cloud center $U^n(m_2)$, randomly generate $M_1=e^{nR_1}$ ``satellite'' codewords $\{X^n(m_1,m_2):  m_1\in \mathcal{M}_1=[M_1]\}$ according to the conditional probability distribution 
\begin{align}\label{def:Xiid}
P (x^n|u^n):=\prod_{i=1}^n P_{X|U}(x_i|u_i).
\end{align}
\item For constant composition random codes, we fix a joint type $P_{UX}$ and randomly and independently generate $M_2=e^{nR_2}$ ``cloud centers'' $\{U^n(m_2) : m_2\in \mathcal{M}_2=[M_2]\}$ under the uniform distribution on the type class $\calT_{P_U}$. For each cloud center $U^n(m_2)$, randomly and independently generate $M_1=e^{nR_1}$ ``satellite'' codewords $\{X^n(m_1,m_2):  m_1\in \mathcal{M}_1=[M_1]\}$ under the uniform distribution on the conditional type class $\calT_{P_{X|U}}(U^n(m_2))$
\end{itemize} 
The two decoders with erasure options are given by {$g_1:\mathcal{Y}^n \to (\mathcal{M}_1 \cup\{\mathrm{e}\})\times (\mathcal{M}_2 \cup\{\mathrm{e}\})$ and 
$g_2:\mathcal{Z}^n \to \mathcal{M}_2\cup\{\mathrm{e}\}$  where $\mathrm{e}$ is the erasure symbol.
\subsection{Definitions of Error Probabilities and Error Exponents}\label{Sec:def_error}
In this paper, {there are essentially twelve error probabilities under consideration: the error probabilities of decoding the pair of messages, the error probabilities of decoding the private message $m_1$ only and the error probabilities of decoding the common message $m_2$ only. For each of these probabilities, there are the total and undetected error probabilities, and each can be computed at any of the terminals.} We focus on six different error probabilities associated to  terminal $\mathcal{Y}$.  We do not derive the total and undetected error probabilities at terminal $\mathcal{Z}$ since the analysis is completely analogous to the analysis of the error and erasure probabilities of the ``cloud centers'' at terminal $\calY$ by replacing $W_{\calY}$ with $W_{\calZ}$. {However, we do compute these exponents numerically in Section~\ref{sec:relation} (see Figure~\ref{fig-ZY}).} Define the disjoint decoding regions according to the decoder $g_1$ as $\mathcal{D}_{m_1m_2}:=\{y^n:g_1(y^n)=(m_1,m_2)\}$. Moreover, let $\{\mathcal{D}_{m_1}:m_1\in\mathcal{M}_1\}$ and $\{\mathcal{D}_{m_2}:m_2\in\mathcal{M}_2\}$ be the disjoint decoding regions associated to   messages $m_1$ and $m_2$ respectively. For terminal $\mathcal{Y}$, define for message  $m_j,j=1,2$ and the message pair $(m_1, m_2)$, the {\em conditional total error}  and {\em undetected error probabilities} as
\begin{align}
&e_j^{\mathrm{t}}(m_1,m_2):= W_{\mathcal{Y}}^n\big(\mathcal{D}^c_{m_j} \, \big|\, x^n(m_1,m_2)\big)\\
&e_j^{\mathrm{u}}(m_1,m_2):= W_{\mathcal{Y}}^n\bigg(\bigcup_{\hat{m}_j\in\mathcal{M}_j\setminus\{m_j\}}\mathcal{D}_{\hat{m}_j} \,\bigg|\, x^n(m_1,m_2)\bigg)\\
&e_Y^{\mathrm{t}}(m_1,m_2):= W_{\mathcal{Y}}^n\left(\mathcal{D}^c_{m_1,m_2}\, \big|\, x^n(m_1,m_2)\right)\\
&e_Y^{\mathrm{u}}(m_1,m_2):= W_{\mathcal{Y}}^n\bigg(\bigcup_{(\hat{m}_1,\hat{m}_2)\neq (m_1,m_2)}\mathcal{D}_{\hat{m}_1\hat{m}_2}\,\bigg|\,  x^n(m_1,m_2)\bigg)  . \label{eqn:eYu}
\end{align}
Then we may define the  {\em average total} and {\em undetected error probabilities} at terminal $\mathcal{Y}$ as follows:
\begin{align}
&e_j^{k}\!:=\!\frac{1}{M_1M_2}\sum_{(m_1,m_2)\in \mathcal{M}_1\times \mathcal{M}_2}e_j^{k}(m_1,m_2), \;\,\, k \in \{ \mathrm{t},\mathrm{u} \}\label{eqn:ejt}\\
&e_Y^{k}\!:=\!\frac{1}{M_1M_2}\sum_{(m_1,m_2)\in \mathcal{M}_1\times \mathcal{M}_2}e_Y^{k}(m_1,m_2), \; k\in\{\mathrm{t},\mathrm{u}\}\label{eqn:eYU} .
\end{align}

Using the Neyman-Pearson theorem,   Forney~\cite{forney1968exponential} obtained the {optimal} trade-off between the average total and undetected error probabilities {for discrete memoryless channels. By following his idea and using a similar argument, we can show that the optimal trade-off between the average total and undetected error probabilities for the ABC is attained by the following decoding regions\footnote{{In the following, the threshold $T$  may take different values depending on whether we are decoding individual messages or the message pair.}}}
\begin{align}
\mathcal{D}^*_{m_j}&:=\bigg\{y^n:\frac{\Pr(y^n|\mathcal{C}_j(m_j))}{\sum_{m'_j\neq m_j}\Pr(y^n|\mathcal{C}_j(m'_j))}\geq e^{nT}\bigg\},   \label{def:DRj}\\
\mathcal{D}^*_{m_1m_2}&:=\bigg\{y^n: \frac{W_{\mathcal{Y}}^n(y^n|x^n(m_1,m_2))}{\sum_{(m'_1,m'_2)\neq (m_1,m_2)}W_{\mathcal{Y}}^n(y^n|x^n(m'_1,m'_2))} \nn\\*
&\qquad \qquad\qquad\qquad \qquad\qquad\qquad \geq  e^{nT}\bigg\},\label{def:DRmp}
\end{align}
where  the distribution of the output $y^n$ conditioned on the subcodebook  $\mathcal{C}_1(m_1) = \{x^n(m_1,m_2):m_2\in\mathcal{M}_2\}$ is 
\begin{align}
\Pr(y^n|\mathcal{C}_1(m_1)) :=\frac{1}{M_2}\sum_{m_2\in \mathcal{M}_2} W_{\mathcal{Y}}^n(y^n|x^n(m_1,m_2))
\end{align}
and similarly for $\Pr(y^n|\mathcal{C}_2(m_2))$.\par

We would like to  find the  {\em exact} error exponents $E^{\mathrm{t}}_j$, $E^{\mathrm{u}}_j$, $E^{\mathrm{t}}_Y$ and $E^{\mathrm{u}}_Y$, $j=1,2$ {with the erasure option, i.e., $T\geq 0$~(we do not consider the list decoding mode, i.e., $T<0$, in this paper)}. These are the exponents associated to the expectation of the error probabilities, where the expectation is taken with respect to  the randomness of the codebook $\mathcal{C}$ which possess the superposition structure as described in Section~\ref{sec:model}. In other words,
\begin{align}\label{def:E1t}
E_1^{\mathrm{t}}(R_1,R_2,T):=\limsup_{n\to \infty}\left[-\frac{1}{n}\ln\mathbb{E}_{\mathcal{C}}[e^{\mathrm{t}}_1]\right],
\end{align}
and similarly for the other exponents $E_1^{\mathrm{u}}, E_Y^{\mathrm{t}}, E_Y^{\mathrm{t}}, E_2^{\mathrm{t}}$, and  $E_2^{\mathrm{u}}$. We show, in fact, that the $\limsup$ in~\eqref{def:E1t} is a limit. These exponents are also called {\em random coding  error exponents}. If these exponents are known exactly, we say that  {\em ensemble-tight} results are established. 
\section{Main Results and Discussions} 
The main result in this paper are stated below in Theorems~\ref{TH:E1} and~\ref{TH:E2}, establishing exact random coding error exponents for the messages $m_j$, $j=1,2$, and the message pair at   terminal~$\mathcal{Y}$, i.e., the random coding exponents corresponding to the probabilities in~\eqref{eqn:ejt}--\eqref{eqn:eYU}. 

Before stating our results, we state a few additional definitions. For a given probability distribution $Q=Q_{UXY}$ on $\mathcal{U}\times\mathcal{X}\times\mathcal{Y}$,  rates $R_1$ and $R_2$, and the fixed   random coding distribution $P=P_{UX}$,  define
\begin{align}
   \beta(Q,R_1) &  := D(Q_{X|U}\|P_{X|U}|Q_U)  \nn\\*
   &\qquad+   I_{Q}(X;Y|U)- R_1\label{def:beta} \\
   \gamma(Q,R_2)&:= D(Q_U \| P_U)+I_Q(U;Y)-R_2\label{def:gamma}\\
   \Phi(Q,R_1,R_2)&:=\big|\gamma(Q,R_2)+|\beta(Q,R_1)|_{+}\big|_{+}\label{def:Phi}\\
   \Delta(Q,R_1,R_2)&:=\big||-\gamma(Q,R_2)|_{+}-\beta(Q,R_1)\big|_{+}.
\end{align}
\subsection{Main Results}
\begin{theorem}\label{TH:E1}
{For \iid random codes,} the error exponents $E_1^{\mathrm{t}}$, $E_1^{\mathrm{u}}$, $E_Y^{\mathrm{t}}$ and $E_Y^{\mathrm{u}}$ are given by\footnote{{In the following analyses and derivations, for ease of notation, we sometimes drop the dependencies of the error exponents (including those in Theorem \ref{TH:E2})  on the parameters $(R_1,R_2,T)$.}}
\begin{align}
 E_1^{\mathrm{t}}(R_1,R_2,T) &=E_Y^{\mathrm{t}}(R_1,R_2,T)\nn\\*
 &=\min\{\Psi_{\mathrm{a}}(R_1,R_2,T),\Psi_{\mathrm{b}}(R_1,T)\}\label{def:TH:E1}\\
  E_1^{\mathrm{u}}(R_1,R_2,T)&=E_Y^{\mathrm{u}}(R_1,R_2,T) \nn\\*
  &=E_1^{\mathrm{t}}(R_1,R_2,T)+T\label{def:TH:E1_u}
\end{align}
where
\begin{align}
&\Psi_{\mathrm{a}}(R_1,R_2,T)\nn\\*
&:=\min_{\hat{Q}_{UXY}}\Big[D(\hat{Q}_{UXY}\|P_{UXY})  \nn\\*
&\quad+\min_{Q_{UX|Y}\in \mathcal{L}_1(\hat{Q}_{XY},R_1,R_2,T)}\Phi(Q_{UX|Y}\hat{Q}_Y,R_1,R_2)\Big]\label{def:Phi_a}\\
&\Psi_{\mathrm{b}}(R_1,T)\nn\\*
&:=\min_{\hat{Q}_{UXY}}\Big[D(\hat{Q}_{UXY}\|P_{UXY})\nn\\*
&\quad+\min_{Q_{X|UY}\in \mathcal{L}_2(\hat{Q}_{UXY},R_1,T)}|\beta(Q_{X|UY}\hat{Q}_{UY},R_1)|_{+}\Big]\label{def:Phi_b}
\end{align}
with $P_{UXY}(u,x,y):= P_{UX}(u,x)W_{\mathcal{Y}}(y|x)$ and  the sets $\mathcal{L}_1$ and $\mathcal{L}_2$ are defined as 
\begin{align}
&\mathcal{L}_1(\hat{Q}_{XY},R_1,R_2,T) \nn\\*
&\!:=\!\Big\{Q_{UX|Y}\!:\!\mathbb{E}_{Q}\ln\frac{1}{W_{\mathcal{Y}}}\!+\!\mathbb{E}_{\hat{Q}}\ln W_{\mathcal{Y}}\!-\!T 
 \!\leq\! \Delta(Q,R_1,R_2)\Big\}\label{def:L1}\\
&\mathcal{L}_2(\hat{Q}_{UXY},R_1,T)\nn\\*
&\!:=\!\Big\{Q_{X|UY}\!:\!\mathbb{E}_{Q}\ln\frac{1}{W_{\mathcal{Y}}}\!+\!\mathbb{E}_{\hat{Q}}\ln W_{\mathcal{Y}}\!-\! T 
 \!\leq\! |-\beta(Q,R_1)\big|_{+}\Big\},\label{def:L2}
\end{align}
where $Q$ in~\eqref{def:L1} is equal to $Q=Q_{UX|Y}\hat{Q}_Y$, $Q$ in~\eqref{def:L2} is equal to $Q=Q_{X|UY}\hat{Q}_{UY}$, and the expectation $\mathbb{E}_{\hat{Q}}\ln W_{\mathcal{Y}}$ can be  explicitly written as $\sum_{u,x,y}\hat{Q}_{UXY}(u,x,y)\ln W_{\mathcal{Y}} (y|x)$.\par
For constant composition random codes, the corresponding error exponents $E_1^{\mathrm{t}}$, $E_1^{\mathrm{u}}$, $E_Y^{\mathrm{t}}$ and $E_Y^{\mathrm{u}}$ can be obtained by adding additional constraints to the optimization problems that define the \iid random coding error exponents above. In particular, all joint distributions $Q_{UXY}$ and $\hatQ_{UXY}$ that appear in~\eqref{def:Phi_a}--\eqref{def:L2}~should satisfy the marginal constraint $Q_{UX}=P_{UX}$. For example, the corresponding exponent $\Psi'_{\rma}$ for constant composition random codes is given by
\begin{align}
&\widetilde{\Psi}_{\mathrm{a}}(R_1,R_2,T)\nn\\*
&:=\min_{\hat{Q}_{UXY}:\hatQ_{UX}=P_{UX}}\Big[D(\hat{Q}_{UXY}\|P_{UXY})\nn\\*
&\quad\  +\min_{Q_{UX|Y}\in \widetilde{\mathcal{L}}_1(\hat{Q}_{XY},R_1,R_2,T)}\Phi(Q_{UX|Y}\hat{Q}_Y,R_1,R_2)\Big]
\end{align}
and the set $\widetilde{\mathcal{L}}_1$ is defined as 
\begin{align}\label{def:L1_1}
&\widetilde{\mathcal{L}}_1(\hat{Q}_{XY},R_1,R_2,T)\nn\\*
&:=\Big\{Q_{UX|Y}:Q_{UX}=P_{UX},\nn\\*
&\quad\quad\quad \mathbb{E}_{Q}\ln\frac{1}{W_{\mathcal{Y}}}+\mathbb{E}_{\hat{Q}}\ln W_{\mathcal{Y}}-T \leq \Delta(Q,R_1,R_2)\Big\}
\end{align}
where $Q$ in~\eqref{def:L1_1} is equal to $Q=Q_{UX|Y}\hat{Q}_Y$ and $Q_{UX}$ in~\eqref{def:L1_1} is the $(\calU\times\calX)$-marginal distribution of $Q$.
\end{theorem}\par
The proof of Theorem~\ref{TH:E1} is provided in Section~\ref{sec:pfTHE1}. It can be shown that there exists a  sequence of (deterministic) codebooks which can simultaneously achieve these following  exponents in Theorems  \ref{TH:E1} and \ref{TH:E2} by using Markov's inequality. (cf. \cite[Proof of Theorem 1]{Hayashi2015asymmetric}).
\begin{theorem}\label{TH:E2}
 For   \iid random codes, the error exponents $E_2^{\mathrm{t}}$ and $E_2^{\mathrm{u}}$ are given by
\begin{align}
\!\! E_2^{\mathrm{t}}(R_1,R_2,T)
&\!=\!\max\{\Psi_{\mathrm{a}}(R_1,R_2,T),\Psi_{\mathrm{c}}(R_1,R_2,T)\},\label{def:ExpM2}\\
\!\! E_2^{\mathrm{u}}(R_1,R_2,T)&\!=\! E_2^{\mathrm{t}}(R_1,R_2,T)+T,
\end{align}
where 
\begin{align}
&\Psi_{\mathrm{c}}(R_1,R_2,T)\nn\\*
&:=\min_{\hat{Q}_{UXY}}\Big[D(\hat{Q}_{UXY}\|P_{UXY})\nn\\
&\quad +\min_{Q_{UX|Y}\in \mathcal{L}_3(\hat{Q}_{UXY},R_1,R_2,T)}\Phi(Q_{UX|Y}\hat{Q}_Y,R_1,R_2)\Big] \label{def:Phi_c}
\end{align}
with
\begin{align}\label{def:L3}
&\mathcal{L}_3(\hat{Q}_{UXY},R_1,R_2,T) \nn\\*
&:=\Big\{Q_{UX|Y}:\mathbb{E}_{Q}\ln\frac{1}{W_{\mathcal{Y}}}+s_0(\hat{Q}_{UY},R_1)-T\nn\\*
&\qquad\qquad\qquad\qquad\qquad\qquad\quad \leq \Delta(Q,R_1,R_2)\Big\}
\end{align}
where $Q$ in~\eqref{def:L3} is equal to $Q=Q_{UX|Y}\hat{Q}_Y$, and 
\begin{align}
&s_0(\hat{Q}_{UY},R_1)\nn\\*
&:= -\min_{\tilde{Q}_{X|UY}:\beta(\tilde{Q},R_1)\leq 0}&\big[\beta(\tilde{Q},R_1)-\mathbb{E}_{\tilde{Q}}\ln W_{\mathcal{Y}}\big]\label{def:s0}
\end{align}
and where $\tilde{Q}$ in~\eqref{def:s0} is equal to $\tilde{Q}=\tilde{Q}_{X|UY}\hat{Q}_{UY}$.\par
 For constant composition random codes, the error exponents $E_2^{\mathrm{t}}$ and $E_2^{\mathrm{u}}$ are given by 
\begin{align}
\!\!\!\! E_2^{\mathrm{t}}(R_1,R_2,T)&\!=\!\max \big\{\widetilde{\Psi}_{\mathrm{a}}(R_1,R_2,T),\widetilde{\Psi}_{\mathrm{c}}(R_1,R_2,T)\big\},\\*
\!\!\!\! E_2^{\mathrm{u}}(R_1,R_2,T)&\!=\! E_2^{\mathrm{t}}(R_1,R_2,T)+T,
\end{align}
where 
\begin{align}
&\widetilde{\Psi}_{\mathrm{c}}(R_1,R_2,T)\nn\\*
&:=\min_{\hat{Q}_{UXY}:\hatQ_{UX}=P_{UX}}\Big[D(\hat{Q}_{UXY}\|P_{UXY})\nn\\*
&\quad +\min_{Q_{UX|Y}\in \widetilde{\mathcal{L}}_3(\hat{Q}_{UXY},R_1,R_2,T)}\Phi(Q_{UX|Y}\hat{Q}_Y,R_1,R_2)\Big] 
\end{align}
with
\begin{align}\label{def:L3_1}
&\widetilde{\mathcal{L}}_3(\hat{Q}_{UXY},R_1,R_2,T)\nn\\*
&:=\Big\{Q_{UX|Y}:Q_{UX}=P_{UX},\nn\\*
&\quad \mathbb{E}_{Q}\ln\frac{1}{W_{\mathcal{Y}}}+\tilde{s}_0(\hat{Q}_{UY},R_1)-T \leq \Delta(Q,R_1,R_2)\Big\}
\end{align}
where $Q$ in~\eqref{def:L3_1} is equal to $Q=Q_{UX|Y}\hat{Q}_Y$, $Q_{UX}$ in~\eqref{def:L3_1} is the $(\calU\times\calX)$-marginal distribution of $Q$ and 
\begin{align}\label{def:s0_1}
&\tilde{s}_0(\hat{Q}_{UY},R_1)\nn\\*
&:= -\min_{\substack{\tilde{Q}_{X|UY}:\beta(\tilde{Q},R_1)\leq 0,\\ \tilQ_{UX}=P_{UX}}}\big[\beta(\tilde{Q},R_1)-\mathbb{E}_{\tilde{Q}}\ln W_{\mathcal{Y}}\big]
\end{align}
and where $\tilde{Q}$ in~\eqref{def:s0_1} is equal to $\tilde{Q}=\tilde{Q}_{X|UY}\hat{Q}_{UY}$ and $\tilQ_{UX}$ in~\eqref{def:s0_1} is the $(\calU\times\calX)$-marginal distribution of $\tilQ$.
\end{theorem}\par
The proof of Theorem~\ref{TH:E2} is provided in Section~\ref{sec:pfTHE2}.

\subsection{Discussion of Main Results}\label{sec:remark}
A few remarks on the theorems above are in order. 

\begin{itemize} 
\item   Eqn.~\eqref{def:TH:E1}  in Theorem~\ref{TH:E1} implies that the optimal decoder for the pair of messages $(m_1, m_2)$ (i.e., $\calD_{m_1m_2}^*$ defined in~\eqref{def:DRmp}) achieves the optimal trade-off between the total and undetected error exponents pertaining to the private message $m_1$. This observation is non-trivial and not immediately obvious. When $\mathcal{Y}$ wishes to decode only the private message $m_1$, the optimal decoder for the {\em pair} of messages $(m_1, m_2)$, called the {\em joint decoder}, declares the message $\hat{m}_1$  of the decoded message pair $(\hat{m}_1, \hat{m}_2)$  is the final output. It is not clear that this decoding strategy is optimal  error exponent-wise. The main difference between the error events for these two decoders is that the user $\mathcal{Y}$ can decode the correct private message $m_1$ but the wrong common message $m_2$. This is an error event for the joint decoder (but not for the one that focuses only on $m_1$). However, Lemma~\ref{lem:remove_dependency} implies that  on the exponential scale, the exponents of the two decoders are the same, i.e., there is no loss in optimality in using the joint decoder for decoding {\em only} message $m_1$. 

\item One of our key technical contributions is Lemma \ref{lem:remove_dependency} (to follow).  This lemma allows us to simplify the calculation of the exponents by disentangling the statistical dependencies between ``satellite codewords'' that share the same cloud center. In particular, when we take into account the fact that the ``cloud centers'' $\mathcal{C}_U'$ (of which there are exponentially many) are random, this lemma allows us to decouple the dependence between  two key random variables 
\begin{equation}
F_1= \sum_{m'_1\in \mathcal{M}_1\setminus\{1\}}\sum_{m'_2\in \mathcal{M}_2\setminus\{1\}} W_{\mathcal{Y}}^n(y^n|X^n(m'_1,m'_2)),
\end{equation}
and 
\begin{equation}
F_4 = \sum_{m'_2\in \mathcal{M}_2\setminus\{1\}}W_{\mathcal{Y}}^n(y^n|X^n(1,m'_2))
\end{equation}

which are on different sides of  a fundamental error probability (see \eqref{eqn:compact_E1} and  \eqref{eqn:simplified_error} in the proof of Theorem~\ref{TH:E1} in Section~\ref{sec:pfTHE1}). In contrast, for the   analysis of the interference channel in~\cite{huleihel2017random} and~\cite{Etkin2010error}, only an  {\em upper bound} of the error probability is sought. This upper bound is not necessarily exponentially tight. On the other hand, the use of Lemma \ref{lem:remove_dependency}  incurs {\em no loss in optimality} on the exponential scale when appropriately combined with Lemma \ref{lem:dependency}. 


\item In an elegant work in~\cite{merhav2014exact}, Merhav showed that for ordinary channel coding, independent random selection of codewords within a given type class together with suboptimal bin index decoding (which is based on  ordinary maximum likelihood decoding), performs as well as optimal bin index decoding in terms of the error exponent achieved. Furthermore, Merhav showed that for constant composition random codes with superposition coding and optimal decoding, the conclusion above no longer holds in general. In this
paper, we show that for \iid and constant composition random codes with superposition coding and erasure decoding, the conclusion holds for the case of  decoding the ``satellite'' codewords. That is the (in general) suboptimal decoding of the ``satellite'' codewords achieves same random coding error exponent as the optimal decoding of the ``satellite'' codewords (see Theorem \ref{TH:E1}). 

\item In Theorem~\ref{TH:E1}, the total error exponent for the private message $m_1$ is the minimum of two exponents $\Psi_{\mathrm{a}}$ and $\Psi_{\mathrm{b}}$. The first exponent $\Psi_{\mathrm{a}}$ intuitively means that the user $\mathcal{Y}$ is in a regime where it decodes the {\em pair} of messages $(\hat{m}_1, \hat{m}_2)$. 
 Loosely speaking, the second   exponent $\Psi_{\mathrm{b}}$ means that   user $\mathcal{Y}$ knows the true common message {$\tilde{m}_2$} (given by a genie), then decodes the ``satellite'' codeword {$X^n(m_1,\tilm_2)$}. In contrast to the single-user DMC case, now every codeword is generated according to a {\em conditional} probability distribution $P_{X|U}$. Thus all codewords are  conditioned on a particular $u^n(\tilm_2)$ sequence rather than being generated according to a marginal distribution $P_X$. This is also reflected in the expression of the inner optimization in~\eqref{def:Phi_b} which is averaged over the random variable $U$ (see definition of $\beta(\cdot)$ in~\eqref{def:beta}).

\item {In this work, while it seems natural, we do not consider the list decoding mode in which $T<0$  due to a couple of technical reasons.  
To ensure that $e^{-n(T+R_1)}$ in \eqref{eqn:dependency_case1} vanishes,  Lemma \ref{lem:dependency} holds on   the condition that $T>-R_1$, rather than the more   general $T<0$. Furthermore, Lemma \ref{lem:remove_dependency}, which is crucial in removing the dependence between two key random variables $F_1$ and $F_4$, requires that  $T\geq 0$ due to the derivation of~\eqref{newlab3}. It appears to the authors that  relaxations of the conditions on $T$ in Lemmas~\ref{lem:dependency} and~\ref{lem:remove_dependency} would be  rather involved and so we defer the consideration of the list decoding mode to future work. }
 

 
\item {It is clear from the closed-form expressions of the exponents in  Theorems~\ref{TH:E1} and~\ref{TH:E2} that the constant composition ones are at least as large as their \iid  counterparts. In Section~\ref{sec:ccvsiid}, we present a numerical example to show that this inequality can be {\em strict}. Furthermore, if the broadcast channel is degraded in favor of $\calY$ (i.e., $X\markov Y\markov Z$ forms a Markov chain in this order), the error exponents at $\calZ$ are smaller than that at $\calY$. We also verify this numerically in Section~\ref{sec:relation}.  } 

\item  Finally, for the case in which user $\mathcal{Y}$ wishes to decode the common message $m_2$, the intuition gleaned from Theorem~\ref{TH:E2} is that if the decoding is not correct, both   events $\{F'_1\geq f_3 e^{-nT}\}$ and  $\{F'_1\geq F_2 e^{-nT}\}$ should occur  (see~\eqref{eqn:E2_compact}). This also means that $\calY$ can  take one of two actions. First, decode the {\em true transmitted codeword} $X^n(m_1,m_2)$ to identify $m_2$ when the complement of the first event (i.e., $\{F'_1\leq f_3 e^{-nT}\}$)  occurs; this corresponds to the exponent $\Psi_\rma$. Second it can decode the {\em sub-codebook for the common message $\mathcal{C}'_2(m_2):=\{X^n(m_1,m_2):m_1\in [M_1]\setminus \{1\}\}$} to identify $m_2$ when the second event $\{F'_1\leq F_2 e^{-nT}\}$ occurs; this corresponds to $\Psi_\rmc$.  This explains the maximization in the first expression in~\eqref{def:ExpM2}. When $R_1$ is large, the term $\Psi_{\mathrm{c}}$ in~\eqref{def:Phi_c} of Theorem~\ref{TH:E2} implies that   $\mathcal{Y}$ is more likely than not to decode the ``cloud center'' $U^n(m_2)$ according to the ``test channel'' $W_{Y|U}(y|u):= \sum_{x} W_{\mathcal{Y}}(y|x)P_{X|U}(x|u)$. This corresponds to the second decoding strategy, i.e., decoding the entire sub-codebook $\mathcal{C}'_2(m_2)$  indexed by $m_2$. Also see Remark~\ref{rmk:two_max} to follow.

\end{itemize}

\section{Evaluating the Exponents via Convex Optimization}
{In this section, we first consider \iid random codes.} To evaluate $E^{\mathrm{t}}_1$ in Theorem~\ref{TH:E1}, we need to devise an efficient numerical procedure to  solve the   minimization problems $\Psi_{\mathrm{a}}$ and $\Psi_{\mathrm{b}}$. As will be shown below, these  problems can be solved efficiently even though they are not convex.\par
For the second term $\Psi_{\mathrm{b}}$ in (\ref{def:Phi_b}), we can split the feasible region of the inner minimization, i.e., $\mathcal{L}_2(\hat{Q}_{UXY},R_1,T)$ (see~\eqref{def:L2}), into two closed sets, namely  $\mathcal{L}_{21}(\hat{Q}_{UXY}):=\mathcal{L}_2(\hat{Q}_{UXY},R_1,T)\cap  \mathcal{B}_1(\hat{Q}_{UY},R_1)$ and $\mathcal{L}_{22}(\hat{Q}_{UXY}):=\mathcal{L}_2(\hat{Q}_{UXY},R_1,T)\cap  \mathcal{B}_2(\hat{Q}_{UY},R_1)$, where
\begin{align}
 \mathcal{B}_1(\hat{Q}_{UY},R_1)&:=\{Q_{X|UY}:\beta(Q_{X|UY}\hat{Q}_{UY},R_1)\geq 0\}\\
 \mathcal{B}_2(\hat{Q}_{UY},R_1)&:=\{Q_{X|UY}:\beta(Q_{X|UY}\hat{Q}_{UY},R_1)\leq 0\}.
\end{align}
We denote the corresponding minimization problems pertaining to $\Psi_{\mathrm{b}}$ in~\eqref{def:Phi_b} (and~\eqref{def:L2}) in which the function  $|\cdot|_{+}$ is inactive or active as $\Psi_{\mathrm{b}1}$ and $\Psi_{\mathrm{b}2}$, respectively, i.e.,
\begin{align}
\Psi_{\mathrm{b}1}&:=\min_{\hat{Q}_{UXY}}\Big[D(\hat{Q}_{UXY}\|P_{UXY})\nn\\*
&\qquad+\min_{Q_{X|UY}\in \mathcal{L}_{21}(\hat{Q}_{UXY})}\beta(Q_{X|UY}\hat{Q}_{UY},R_1)\Big]\\
\Psi_{\mathrm{b}2}&:=\min_{\hat{Q}_{UXY}:\mathcal{L}_{22}(\hat{Q}_{UXY})\neq \emptyset}D(\hat{Q}_{UXY}\|P_{UXY}),
\end{align}
where  the sets $\mathcal{L}_{21}$ and $\mathcal{L}_{22}$ are defined as
\begin{align}
&\mathcal{L}_{21}(\hat{Q}_{UXY})\nn\\*
&:=\Big\{Q_{X|UY}:\mathbb{E}_{Q}\ln\frac{1}{W_{\mathcal{Y}}}+\mathbb{E}_{\hat{Q}}\ln W_{\mathcal{Y}}-T \leq 0,\nn\\*
&\qquad\qquad\qquad\qquad\qquad\qquad\qquad\quad \beta(Q,R_1)\geq 0\Big\},\label{ExpL2a}\\
&\mathcal{L}_{22}(\hat{Q}_{UXY})\nn\\*
&:=\Big\{Q_{X|UY}:\mathbb{E}_{Q}\ln\frac{1}{W_{\mathcal{Y}}}+\mathbb{E}_{\hat{Q}}\ln W_{\mathcal{Y}}-T+\beta(Q,R_1) \leq 0,\nn\\*
&\qquad\qquad\qquad\qquad\qquad\qquad\qquad\quad \beta(Q,R_1)\leq 0\Big\}\label{ExpL2b} ,
\end{align}
and where $Q$ in~\eqref{ExpL2a} and~\eqref{ExpL2b} is equal to $Q=Q_{X|UY}\hat{Q}_{UY}$. We thus have $\Psi_{\mathrm{b}}=\min\{\Psi_{\mathrm{b}1},\Psi_{\mathrm{b}2}\}$.\par 
As the minimization problem $\Psi_{\mathrm{b}2}$ is convex, it can be solved efficiently. However $\Psi_{\mathrm{b}1}$ is non-convex due to the non-convex constraint $\beta(Q_{X|UY}\hat{Q}_{UY})\geq 0$ in the inner optimization.\footnote{In this section, we drop the dependences of $\beta(\cdot)$ and $\gamma(\cdot)$ on the rates $R_1$ and $R_2$}  For the inner optimization, if we remove  this constraint in $\mathcal{L}_{21}$, the modified problem is
\begin{align}
\Psi'_{\mathrm{b}1}&:=\min_{\hat{Q}_{UXY}}\Big[D(\hat{Q}_{UXY}\|P_{UXY})\nn\\*
&\qquad +\min_{Q_{X|UY}\in \mathcal{L}'_{21}(\hat{Q}_{UXY})}\beta(Q_{X|UY}\hat{Q}_{UY})\Big],
\end{align}
where
\begin{align}
&\mathcal{L}'_{21}(\hat{Q}_{UXY})\nn\\*
&:=\Big\{Q_{X|UY}:\mathbb{E}_{Q}\ln\frac{1}{W_{\mathcal{Y}}}+\mathbb{E}_{\hat{Q}}\ln W_{\mathcal{Y}}-T \leq 0\Big\},
\end{align}
is convex and can be solved efficiently. Furthermore, we have the following proposition.\par
\begin{proposition}\label{prop:op1}
For the optimization problem $\Psi_{\mathrm{b}1}$, if the   optimal solution to the inner optimization of the modified problem $\Psi'_{\mathrm{b}1}$ is not feasible for the original problem $\Psi_{\mathrm{b}1}$, i.e., $\beta(Q_{X|UY}\hat{Q}_{UY})<0$, then there exists an optimal solution to the original inner optimization problem that satisfies $\beta(Q_{X|UY}\hat{Q}_{UY})=0$. Moreover, in this case, the optimal value of $\Psi_{\mathrm{b}}$ is equal that for  $\Psi_{\mathrm{b}2}$ (i.e., $\Psi_{\mathrm{b}2}$ is active in the minimum that defines $\Psi_{\mathrm{b}}$).
\end{proposition}
\begin{IEEEproof}[Proof of Proposition~\ref{prop:op1}]
{See Appendix~\ref{pf:prop_op1}.}
\end{IEEEproof}
In summary, we can solve the non-convex optimization problem $\Psi_{\mathrm{b}}$ by solving two convex problems $\Psi_{\mathrm{b}2}$ and $\Psi'_{\mathrm{b}1}$, i.e.,
\begin{align}
\Psi_{\mathrm{b}}=\min\{(\Psi'_{\mathrm{b}1})^{+},\Psi_{\mathrm{b}2}\},
\end{align}
where the superscript ``$+$'' of $(\Psi'_{\mathrm{b}1})^{+}$ means the value of $(\Psi'_{\mathrm{b}1})^{+}$ is {\em active} in the minimization if the optimal solution $(\hat{Q}^*,Q^*)$ is also feasible for the original optimization $\Psi_{\mathrm{b}1}$, i.e., $\beta(Q^*_{X|UY}\hat{Q}^*_{UY})\geq0$. In other words,
\begin{align}\label{defact}
\Psi_{\mathrm{b}}=\left\{
\begin{aligned}
&\min\{\Psi'_{\mathrm{b}1},\Psi_{\mathrm{b}2}\}\quad &&\beta(Q^*_{X|UY}\hat{Q}^*_{UY})\geq0\\
&\Psi_{\mathrm{b}2} &&\text{else}
\end{aligned}
\right. .
\end{align}
 Consequently, $\Psi_{\mathrm{b}}$ can be solved efficiently.\par

For $\Psi_{\mathrm{a}}$ in~\eqref{def:Phi_a}, let 
\begin{align}\label{def:Omega}
&\Omega(Q_{UX|Y}\hat{Q}_Y)\nn\\*
&:=\mathbb{E}_{Q_{UX|Y}\hat{Q}_Y}\ln\frac{1}{W_{\mathcal{Y}}}+\mathbb{E}_{\hat{Q}}\ln W_{\mathcal{Y}}-T,
\end{align}
then similarly, we can partition the feasible region of the inner minimization into four parts and denote the corresponding inner optimization problems as follows:
\begin{enumerate}
\item If $\gamma(Q)\geq 0$ and $\beta(Q)\geq 0$, then 
\begin{align}
\Phi_{\mathrm{a}1}^*:=\min_Q\gamma(Q)+\beta(Q),\ \ \mbox{such that}\ \ \Omega(Q)\leq 0.
\end{align}
\item If $\gamma(Q)\geq 0$ and $\beta(Q)\leq 0$, then
\begin{align}
\Phi_{\mathrm{a}2}^*:=\min_Q\gamma(Q), \ \ \mbox{such that} \ \ \Omega(Q)+\beta(Q) \leq 0.
\end{align}
\item If $\gamma(Q)\leq 0$ and $\gamma(Q)+\beta(Q)\geq 0$, then
\begin{align}
\Phi_{\mathrm{a}3}^*:=\min_Q\gamma(Q)+\beta(Q),\ \ \mbox{such that}\ \ \Omega(Q)\leq 0.
\end{align}
\item If $\gamma(Q)\leq 0$ and $\gamma(Q)+\beta(Q)\leq 0$, then
\begin{align}
\Phi_{\mathrm{a}4}^*:= 0 \  \ \mbox{such that}\ \ \Omega(Q)+\gamma(Q)+\beta(Q)\leq 0.
\end{align}
\end{enumerate}
where $Q$ in the above definitions is equal to $Q=Q_{UX|Y}\hat{Q}_Y$ 
(compare the above to the definition of the optimization problem $\Psi_{\mathrm{a}}$ in~\eqref{def:Phi_a}). Thus we have,
\begin{align}
\Psi_{\mathrm{a}}=\min_{\hat{Q}_{UXY}}\big[D(\hat{Q}_{UXY}\|P_{UXY})+\min_{i\in[4]}\{\Phi^*_{\mathrm{a}i}({\hat{Q}_{XY}})\}\big]\label{opt1} .
\end{align} 
We can rewrite the objective functions of $\Phi_{\mathrm{a}1}^*$  and $\Phi_{\mathrm{a}3}^*$  as follows
\begin{align}
&\min_{Q_{UX|Y}}\gamma(Q_{UX|Y}\hat{Q}_Y)+\beta(Q_{UX|Y}\hat{Q}_Y)\nonumber\\*
&=\min_{Q_{U|Y}}[\gamma(Q_{U|Y}\hat{Q}_Y)+\min_{Q_{X|UY}}\beta(Q_{X|UY}Q_{U|Y}\hat{Q}_Y)]
\end{align}
where the notation $\gamma(Q_{U|Y}\hat{Q}_Y)$ is consistent due to the fact that the function $\gamma(Q,R_2)$ (see~\eqref{def:gamma}) only depends on the marginal distribution $Q_{UY}$.
Therefore, by using a similar argument as that for  $\Psi_{\mathrm{b}1}$ above, we can remove the non-convex constraint $\beta(Q)\geq 0$ in $\Phi_{\mathrm{a}1}^*$ due to $\Phi_{\mathrm{a}2}^*$. We can  also remove the non-convex constraint $\gamma(Q)+\beta(Q)\geq 0$ in $\Phi_{\mathrm{a}3}^*$ due to $\Phi_{\mathrm{a}4}^*$. Denote these two modified optimizations as $\Phi_{\mathrm{a}1}'$ and $\Phi_{\mathrm{a}3}'$, respectively. We can merge these  two modified optimizations $\Phi_{\mathrm{a}1}'$ and $\Phi_{\mathrm{a}3}'$ into a new convex optimization problem $\Phi_{\mathrm{a}5}^*$  i.e., 
\begin{align}
\Phi_{\mathrm{a}5}^*:=\min\gamma(Q)+\beta(Q) \quad \mbox{such that} \quad \Omega(Q)\leq 0.
\end{align}
We now state and prove a proposition that simplifies the calculation of~\eqref{opt1}.
\begin{proposition}\label{prop:op2}
For the inner minimization problem in~\eqref{opt1}, i.e., $\min_{i\in[4]}\{\Phi^*_{\mathrm{a}i}(\hat{Q}_{XY})\}$, without loss of optimality, we can replace $\Phi_{\mathrm{a}1}^*$ and $\Phi_{\mathrm{a}3}^*$ with the new convex optimization problem $\Phi_{\mathrm{a}5}^*$.
\end{proposition}
\begin{IEEEproof}[Proof of Proposition~\ref{prop:op2}]
{See Appendix~\ref{pf:prop_op2}.}
\end{IEEEproof}
For the second term $\Psi_{\mathrm{a}2}$, we can also remove the non-convex constraint $\gamma(Q)\geq 0$ in $\Phi_{\mathrm{a}2}^*$ due to $\Phi_{\mathrm{a}4}^*$. Therefore, we can   solve the minimization problem $\Psi_{\mathrm{a}}$ in~\eqref{def:Phi_a} efficiently, as the remaining case $\Phi_{\mathrm{a}4}^*$ is a convex minimization problem.

Similarly to the   above, we can also efficiently calculate  $E_2^{\mathrm{t}}$ in Theorem~\ref{TH:E2}  as $s_0(\hat{Q},R_1)$ is a convex minimization problem. \par
{Finally, for constant composition random codes, since the additional marginal constraints are linear, the transformed optimization problems remain convex and can be solved efficiently as we show in Section~\ref{sec:ccvsiid}.}
\section{Numerical Evaluations} \label{sec:num}
 
In this section, we present  numerical examples to illustrate the following.
\begin{itemize}
\item The  behavior of the \iid   exponents in Theorems~\ref{TH:E1} and~\ref{TH:E2}; 
\item The comparison between the constant composition and the \iid error exponents in Theorem~\ref{TH:E1}; 
\item The   comparison between the \iid error exponents for message $m_2$ at terminals $\calY$ and $\calZ$. 
\end{itemize}
We consider   binary symmetric channels (BSCs): $Y=X\oplus V_1$ and $Z=X\oplus V_2$, where $X,Y,V_1,V_2\in \{0,1\}$, $V_1\sim \text{Bern}(p_1)$ and $V_2\sim \text{Bern}(p_2)$. Let $U$ be binary as well and $U\sim \text{Bern}(0.5)$. Also, let $X=U\oplus V_3$, where $V_3\in \{0,1\}$ and $V_3\sim \text{Bern}(q)$. In this example, we fix $p_1=0.2$, $p_2=0.25$ and $q=0.1$, and all the rates are in nats. 

All the Matlab\textsuperscript{\textregistered}  code to reproduce Figures~\ref{fig-R2}--\ref{fig-ZY} can be found at \url{https://www.ece.nus.edu.sg/stfpage/vtan/isit18.zip}.
\begin{figure}[t]
\centerline{\includegraphics[width=3.25in]{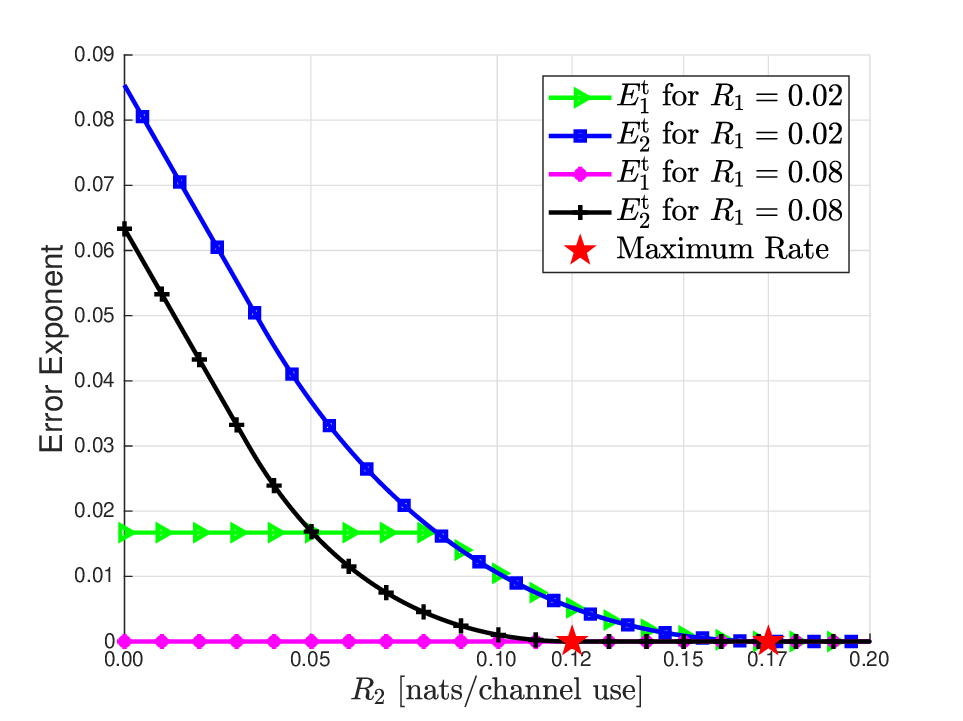}}
\caption{Total error exponents $E_1^{\mathrm{t}}$ and $E_2^{\mathrm{t}}$ as a function of $R_2$  for two different values of $R_1$ and  where the threshold $T=0$.}
\label{fig-R2}
\end{figure}
\begin{figure}[t]
\centerline{\includegraphics[width=3.25in]{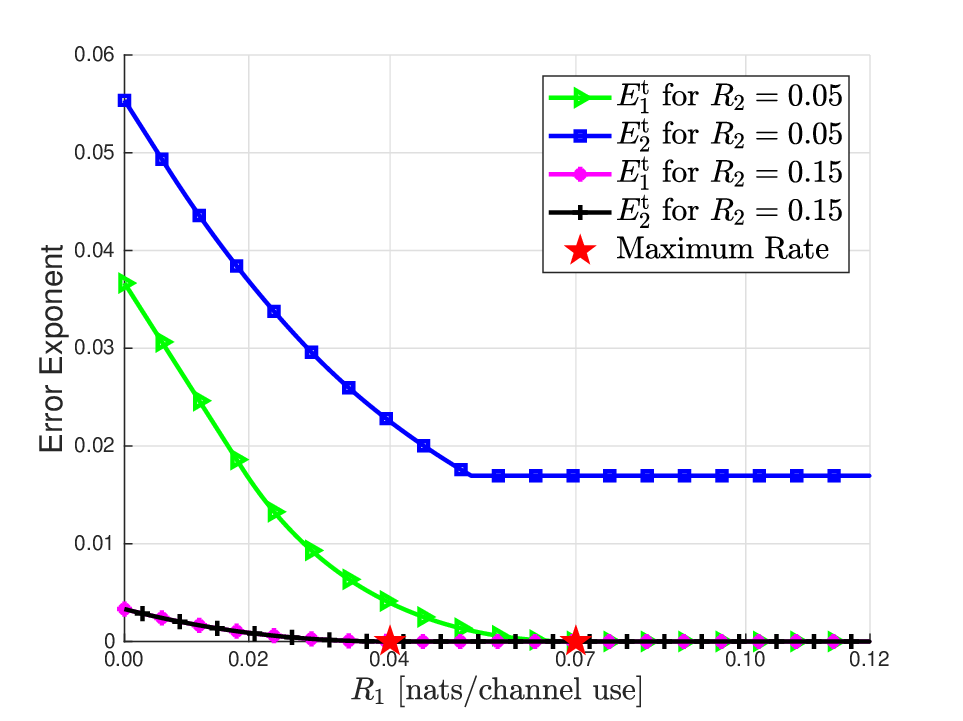}}
\caption{Total error exponents $E_1^{\mathrm{t}}$ and $E_2^{\mathrm{t}}$ as a function of $R_1$ for two different values of $R_2$ and where the threshold $T=0$.}
\label{fig-R1}
\end{figure}
\begin{figure}[t]
\centerline{\includegraphics[width=3.25in]{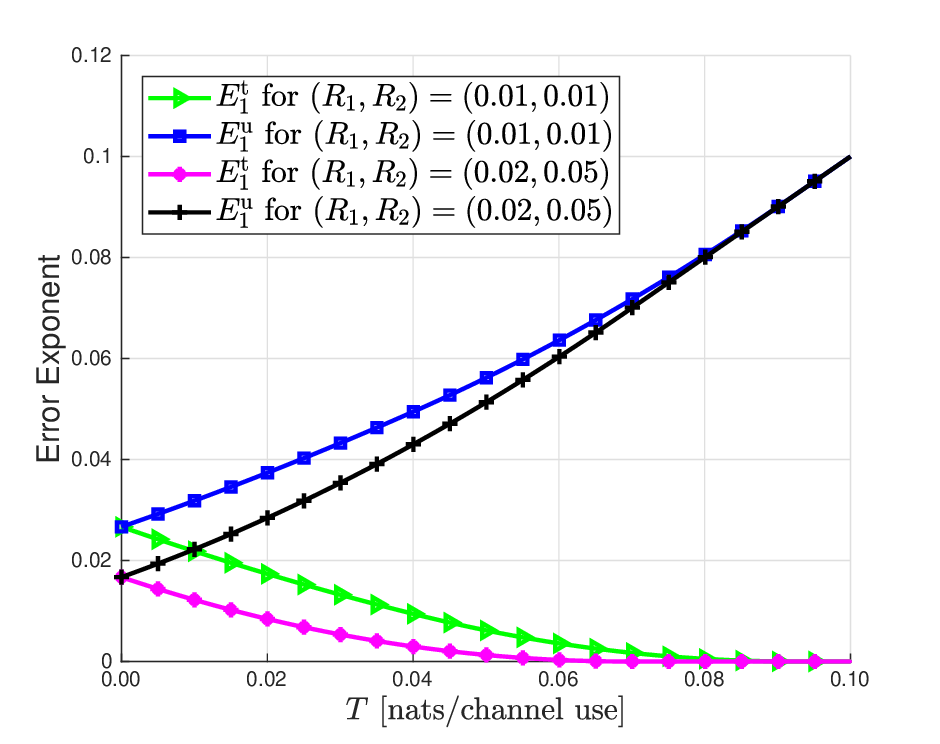}}
\caption{Total error exponent $E^{\mathrm{t}}_1$ and undetected error exponent $E^{\mathrm{u}}_1$ for message $m_1$ as a function of $T$ for two different   pairs of $(R_1,R_2)$.}
\label{fig-TM1}
\end{figure}
\begin{figure}[t]
\centerline{\includegraphics[width=3.25in]{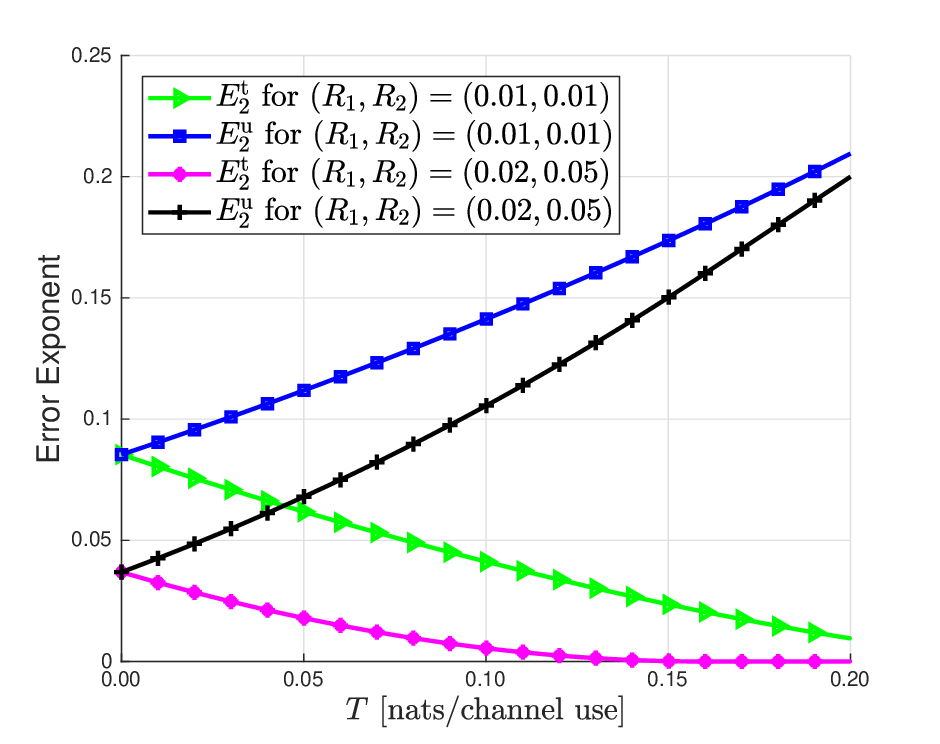}}
\caption{Total error exponent $E^{\mathrm{t}}_2$ and undetected error exponent $E^{\mathrm{u}}_2$ for message $m_2$ as a function of $T$ for two different   pairs of $(R_1,R_2)$.}
\label{fig-TM2}
\end{figure}

\subsection{Behavior of \iid Exponents} \label{sec:iid_exp}
Firstly, we consider the \iid case in which $T=0$.  We obtain a three-dimensional   exponent-rate region $(E^{\mathrm{t}}_j,R_1,R_2)$ for decoding $(m_1, m_2)$. To obtain a two-dimensional plot, we consider projections:   Fix one rate and vary the other rate and plot the error exponent $E^{\mathrm{t}}_j$, $j=1,2$. Figure~\ref{fig-R2} shows one projection for $R_1=0.02$ and $R_1=0.08$ nats/channel use. For message $m_1$, the range of $R_2$ for which $E^{\mathrm{t}}_1>0$ (i.e., $R_2<0.17$ for $R_1=0.02$ and $R_2=0$ for $R_1=0.08$) coincides with that for the set of achievable rate pairs $(R_1,R_2)$ corresponding to our choice of input distribution $P_{UX}$ for decoding  only message $m_1$, namely\footnote{The rate region in \eqref{capacityR1} and \eqref{capacityR2} can be obtained by applying the packing lemma in~\cite[Lemma 3]{el2011network}. Also see \cite[Sec.\ 5.3.1]{el2011network} for a similar analysis of the superposition coding inner bound.} 
\begin{align}\label{capacityR1}
&\{R_1\leq I(X;Y|U)=0.07\}   \nn\\*
&\qquad\, \bigcap\, \{ R_1+R_2\leq I(X;Y)=0.19\}.
\end{align}
Moreover, we see that $E^{\mathrm{t}}_1$ for a fixed $R_1$ is horizontal 
for $R_2$ below a critical value and curved for $R_2$ above this   value. For message $m_1$, the range of $R_2$ for which $E^{\mathrm{t}}_2>0$ (i.e., $R_2<0.17$ for $R_1=0.02$ and $R_2<0.12$ for $R_1=0.08$) coincides that for the set of achievable rate pairs $(R_1,R_2)$ corresponding to our choice of input distribution $P_{UX}$ for  decoding only   $m_2$, i.e., 
\begin{align}\label{capacityR2}
&\{ R_2\leq I(U;Y)=0.12\} \nn\\*
&\qquad\, \bigcup\, \{ R_1+R_2\leq I(X;Y)=0.19\}.
\end{align} 
Figure~\ref{fig-R1} shows the other projection for $R_2=0.05$ and $R_2=0.15$ nats/channel use. It also can be checked that the range of $R_1$ for both messages $(m_1,m_2)$  coincides with~\eqref{capacityR1}. When $R_2=0.05\leq I(U;Y)$, we see the curve of $E^{\mathrm{t}}_2$ rapidly decreases for $R_1$ below a critical value and remains horizontal for $R_1$ above the critical value. This is because when $R_2=0.05\leq I(U;Y)$, the rate pair $(R_1,R_2)$ is always achievable, i.e., $(R_1, R_2)$ belongs to the region defined in~\eqref{capacityR2}. When $R_2=0.15\geq I(U;Y)$, we observe that the two error exponents $E^{\mathrm{t}}_1$ and $E^{\mathrm{t}}_2$ are equal.

Figures~\ref{fig-TM1} and~\ref{fig-TM2} illustrate the optimal trade-off between the \iid total error exponent and the \iid undetected error exponent as function of $T$ for two different   pairs of $(R_1,R_2)$. We observe that for both messages, the total error exponent decreases and the undetected error exponent increases when the threshold $T$ increases. We also observe that the smallest threshold $T$ for which the total error exponent is zero depends on the rate pair $(R_1,R_2)$ and   decreases as either rate increases.

\begin{figure}[t]
\centerline{\includegraphics[width=3.25in]{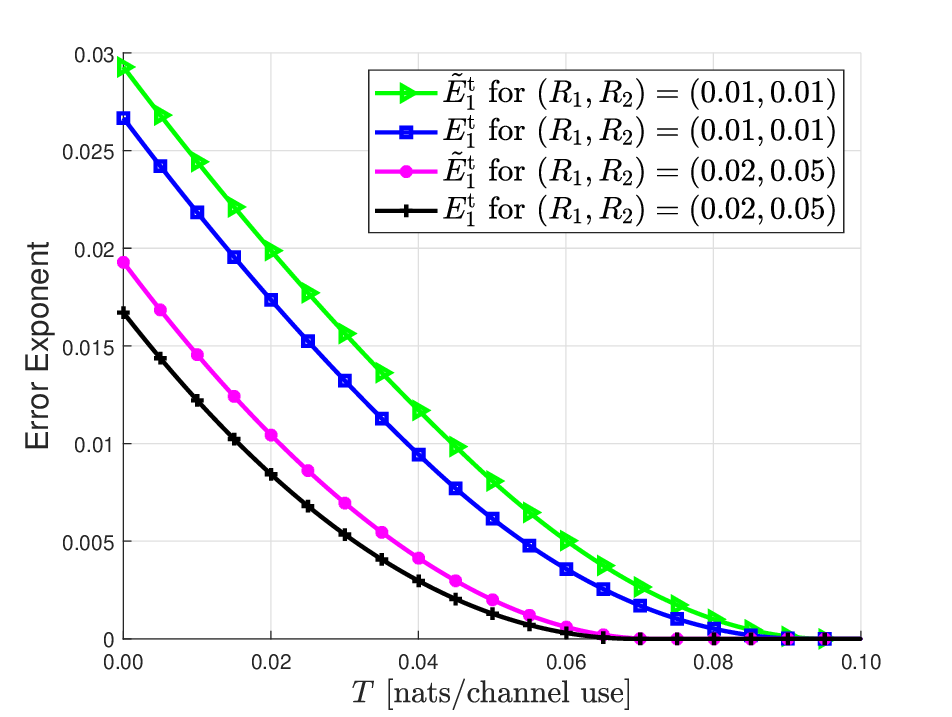}}
\caption{Constant composition and \iid total error exponents $\tilE^{\mathrm{t}}_1$ and $E^{\mathrm{t}}_1$ for message $m_1$ as a function of $T$ for two different   pairs of $(R_1,R_2)$.}
\label{fig-CCIID}
\end{figure}

\subsection{Gain of Constant Composition Exponents over \iid Ones}\label{sec:ccvsiid}
{We now demonstrate the gain of the constant composition exponents over the \iid ones in Theorem~\ref{TH:E1}. Denote the constant composition and \iid total error exponents for  $m_1$ as $\tilE^{\mathrm{t}}_1$ and $E^{\mathrm{t}}_1$ respectively. For the example of BSCs described at the start of this section,  Figure~\ref{fig-CCIID} displays these exponents   as functions of $T$ for two different   pairs of $(R_1,R_2)$. We observe that the constant composition exponents are strictly larger than their \iid counterparts. }
\subsection{Comparison of \iid Exponents at Two Terminals} \label{sec:relation}
{Finally, we consider the relationship between the exponents at the two terminals. We denote the \iid total and undetected error exponent for $m_2$ at terminal $\calZ$ as $E^\rmt_{2,\calZ}$ and $E^\rmu_{2,\calZ}$ respectively. Figure~\ref{fig-ZY} compares the optimal trade-off between these exponents as functions of $T$ for terminals $\calY$ and $\calZ$. We observe that similar to the standard decoding, if the channel quality is worse, this leads to a smaller exponent for the decoding with erasure option (and vice versa).  }

\begin{figure}[t]
\centerline{\includegraphics[width=3.25in]{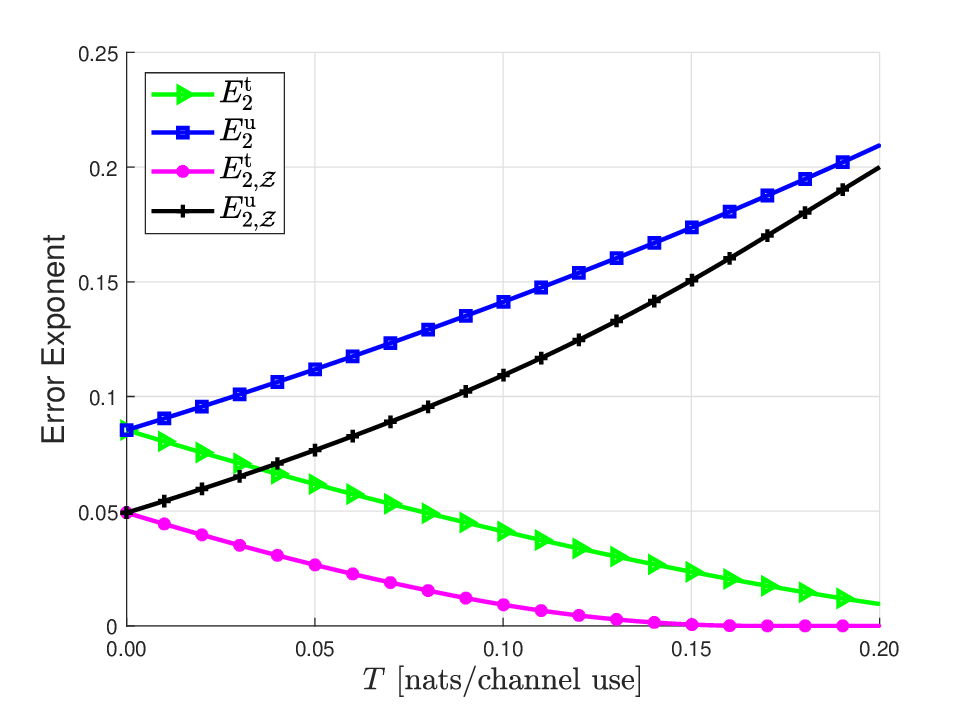}}
\caption{Total and undetected error exponents $E^{\mathrm{u}}_2$ and $E^{\mathrm{t}}_2$ at terminal $\calY$ and $E^{\mathrm{u}}_{2,\calZ}$ and $E^{\mathrm{t}}_{2,\calZ}$ at terminal $\calZ$ for message $m_2$ as a function of $T$ for a given rate pair $(R_1,R_2)=(0.01,0.01)$.}
\label{fig-ZY}
\end{figure}

\section{Proof of  Theorem~\ref{TH:E1}}\label{sec:pfTHE1}
\begin{IEEEproof}[Proof of  Theorem~\ref{TH:E1}] 
 Firstly, we consider i.i.d.\ random codes. At the end of the proof, we describe how to extend the analysis to constant composition codes.  Assume, without loss of generality, that the true transmitted message pair is $(m_1,m_2)=(1,1)$. Denote the  random sub-codebook $\{U^n(m'_2): m'_2\in\mathcal{M}_2\setminus \{1\} \}$ as $\mathcal{C}'_U$, and the (total) error event $\mathcal{E}_1$ as
\begin{align}\label{errorevent1}
\mathcal{E}_1\!:=\! \bigg\{\sum_{m'_1\neq 1} \Pr(Y^n|\mathcal{C}_1(m'_1))  \!>\!  \Pr(Y^n|\mathcal{C}_1(1))e^{-nT}\bigg\}.
\end{align}
Given the optimal decoding region $\mathcal{D}_{m_1}^*$ in~\eqref{def:DRj}, by using the law of total probability, the average total error probability for message $m_1=1$ is 
\begin{align}\label{eqn:error1t}
&\mathbb{E}_{\mathcal{C}}[e_1^{\mathrm{t}}(1,1)]=\mathbb{E}_{(U^n(1),X^n(1,1),Y^n)} \nn\\*
&\qquad\qquad \Big[\mathbb{E}_{\mathcal{C}'_U} \big[\Pr\{\mathcal{E}_1|(U^n(1),X^n(1,1),Y^n),\mathcal{C}'_U\} \big] \Big],
\end{align}

Next, we calculate  the error probability given $(U^n(1),X^n(1,1),Y^n)=(u^n,x^n,y^n)$ with joint type $\hat{Q}_{UXY}$ and  the sub-codebook $\calC'_U=c'_U$. For brevity, define the quantities
\begin{align}
F_1& := \sum_{m'_1\in \mathcal{M}_1\setminus\{1\}}\sum_{m'_2\in \mathcal{M}_2\setminus\{1\}} W_{\mathcal{Y}}^n(y^n|X^n(m'_1,m'_2))\label{def:F1}\\
F_2& := \sum_{m'_1\in \mathcal{M}_1\setminus\{1\}}W_{\mathcal{Y}}^n(y^n|X^n(m'_1,1))\label{def:F2}\\
f_3& := W_{\mathcal{Y}}^n(y^n|x^n) \label{def:f3}\\
F_4& := \sum_{m'_2\in \mathcal{M}_2\setminus\{1\}}W_{\mathcal{Y}}^n(y^n|X^n(1,m'_2))\label{def:F4} .
\end{align}
Note that $f_3$ is a deterministic quantity given $(X^n(1,1),Y^n)=(x^n,y^n)$ while the others are random.   These definitions allow us to express $\Pr\{\mathcal{E}_1\}$ compactly as\footnote{{In the following analysis, for ease of notation, we drop the conditioning events $\{U^n(1),X^n(1,1),Y^n)=(u^n,x^n,y^n)\}$ and $\{\calC'_U=c'_U\}$ when there is no possibility of  confusion.}}
\begin{align}
\Pr\{\mathcal{E}_1\}=\Pr\left\{ F_1+F_2 >(f_3+F_4)\cdot e^{-nT}\right\} \label{eqn:compact_E1} .
\end{align}
Let $Q_{U|Y}$, $Q_{UX|Y}$ and $Q_{X|UY}$ be conditional types such that $Q_{U|Y}\hatQ_Y$, $Q_{UX|Y}\hatQ_Y$ and $Q_{X|UY}\hatQ_{UY}$ are joint types defined on $\mathcal{U}\times\mathcal{Y}$, $\calU\times\calX\times\calY$ and $\calU\times\calX\times\calY$, respectively. Define the following quantities
\begin{align}
\Lambda(Q_{UY},\calC'_U)&:=\big|\big\{U^n(m_2):m_2\in\calM_2\setminus \{1\}, \nn\\*
&\qquad (U^n(m_2),y^n)\in\calT_{Q_{UY}}\big\}\big|,\label{def:U_num}\\
N_{m_1}(Q_{UXY})&:=\big|\big\{X^n(m_1,m_2):m_2\in\calM_2\setminus \{1\}, \nn\\*
&\qquad (u^n(m_2),X^n(m_1,m_2),y^n)\in\calT_{Q_{UXY}}\big\}\big|\nn\\*
& \qquad\qquad \qquad  \mbox{for all}\;\; m_1\in\calM_1,\label{def:XU_num}\\
N(Q_{UXY})&:=\big|\big\{X^n(m_1,1):m_1\in\calM_1\setminus \{1\}, \nn\\*
&\qquad (X^n(m_1,1),u^n,y^n) \in\calT_{Q_{UXY}}\big\}\big|.\label{def:X_num}
\end{align}
 which represent the  {number} of codewords $X^n(m_1,m_2)$ (resp.\ $U^n(m_2)$) whose joint types with the corresponding ``cloud centers'' $u^n(m_2)$ 
and the received sequence $y^n$~(resp.\ and only the received sequence $y^n$) are $Q_{UXY}$ (resp.\ $Q_{UY}$), i.e.,

Note that $\Lambda(Q_{U|Y}\hatQ_{Y},c'_U)$ is a deterministic quantity given $Y^n=y^n$ and a  fixed $ \calC'_U=c'_U$. However, if we take into account the fact that $\mathcal{C}_U'$ is a collection of  random variables, then $\Lambda(Q_{UY},\calC'_U)$ is a random variable given  $Y^n=y^n$.\par 
Now, recall the \iid and constant composition random codebook generation procedures (see Section \ref{sec:model}) and the definitions of $\beta(Q_{UXY},R_1)$ and $\gamma(Q_{UXY},R_2)$ (see \eqref{def:beta} and~\eqref{def:gamma}). Then, $\Lambda(Q_{U|Y}\hatQ_{Y},\calC'_U)$, $N_i(Q_{UX|Y}\hatQ_{Y})$, $i\in \calM_1$ and $N(Q_{X|UY}\hatQ_{UY})$ possess the following properties:

\begin{fact}\label{fact}
\begin{enumerate}
\item For a given $Q_{U|Y}$, $\Lambda(Q_{U|Y}\hatQ_{Y},\calC'_U)$ is a binomial random variable with $(e^{nR_2}-1)$ trials and ``success'' probability\footnote{ In Fact \ref{fact}, $P_{U^n,X^n}$ (in \eqref{def:succ_u}, \eqref{def:succ_x} and \eqref{def:succ_xc})  either denotes the \iid distribution defined in \eqref{def:Uiid} and~\eqref{def:Xiid} or the uniform distribution over the type class $\calT_{P_{UX}}$. See discussion in Section~\ref{sec:model}.}
\begin{align}\label{def:succ_u}
\!\!\frac{\big|\mathcal{T}_{Q_{U|Y}}(y^n)\big|}{\big|\mathcal{T}_{Q_{U}}\big|}\cdot P_{U^n} \big(\mathcal{T}_{Q_{U}}\big)&\doteq\! e^{-n[D(Q_{U}\|P_{U})  +  I_{Q_{UY}}(U;Y)]}\nn\\*
&=e^{-n[\gamma(Q_{UY},R_2)+R_2]},
\end{align}
where $Q_{UY}=Q_{U|Y}\hatQ_Y$. Note that  the notation $\gamma(Q_{UY},R_2)$ is consistent since the function $\gamma(Q,R_2)$ (see~\eqref{def:gamma}) only depends on the marginal distribution $Q_{UY}$. 
\item For a given $Q_{UX|Y}$, { $N_{m_1}(Q_{UX|Y}\hatQ_{Y})$, $m_1\in \calM_1$} are \iid  binomial random variables each with $\Lambda(Q_{U|Y}\hatQ_{Y},c'_U)$ trials and ``success'' probability 
\begin{align}\label{def:succ_x}
&\frac{\big|\mathcal{T}_{Q_{X|UY}}(\tilu^n,y^n)\big|}{\big|\mathcal{T}_{Q_{X|U}}(\tilu^n)\big|}\cdot P_{X^n|U^n} \big(\mathcal{T}_{Q_{X|U}}(\tilu^n)|\tilu^n\big)\nn\\*
&\qquad\doteq e^{-n[D(Q_{X|U}\|P_{X|U}|Q_U)  +  I_{Q_{UXY}}(X;Y|U)]}\nn\\*
&\qquad=e^{-n[\beta(Q_{UXY},R_1)+R_1]},
\end{align}
where $Q_{UXY}=Q_{UX|Y}\hatQ_{Y}$ and $(\tilu^n,y^n)\in \calT_{Q_{UY}}$.
\item For a given $Q_{X|UY}$, $N(Q_{X|UY}\hatQ_{UY})$ is a binomial random variable with $(e^{nR_1}-1)$ trials and ``success'' probability 
\begin{align}\label{def:succ_xc}
&\frac{\big|\mathcal{T}_{Q_{X|UY}}(u^n,y^n)\big|}{\big|\mathcal{T}_{Q_{X|U}}(u^n)\big|}\cdot  P_{X^n|U^n} \big(\mathcal{T}_{Q_{X|U}}(u^n)|u^n\big) \nn\\*
&\qquad\doteq e^{-n[D(Q_{X|U}\|P_{X|U}|\hatQ_U)  +  I_{Q_{UXY}}(X;Y|U)]}\nn\\*
&\qquad=e^{-n[\beta(Q_{UXY},R_1)+R_1]},
\end{align}
where $Q_{UXY}=Q_{X|UY}\hatQ_{UY}$.
\end{enumerate}
\end{fact}
By using a standard large deviations analysis, we obtain the following proposition which is useful to analyze the concentration properties of the random variables defined in \eqref{def:U_num}--\eqref{def:X_num}.
\begin{proposition}\label{prop:LLN}
Suppose $V_i,i=1,\ldots, e^{nr}$, where $r>0$, are \iid Bernoulli random variables with $\mathbb{E}[V_i]=e^{-np}$, 
 where $p>0$. We have
\begin{enumerate}
\item The probability of the event $\{\sum_{i=1}^{e^{nr}} V_i\geq 1\}$ is 
\begin{equation}\label{eqn:prob_geq1}
\Pr\left\{\sum_{i=1}^{e^{nr}} V_i\geq 1\right\}\doteq e^{-n|p-r|_+}.
\end{equation}
\item Let $a=|r-p|_+ +\epsilon\in (0,r)$ where $\epsilon>0$, then the probability of the event $\{\ln\sum_{i=1}^{e^{nr}} V_i\geq na\}$ decays  doubly exponentially, i.e.,
\begin{equation}\label{eqn:prob_geqdouble}
\Pr\left\{\sum_{i=1}^{e^{nr}} V_i\geq e^{na}\right\}\leq \exp\{-e^{na}[n(p+a-r)-1]\}.
\end{equation}
\item Assume $r>p$ and let $a=r-p-\epsilon>0$ where $\epsilon>0$, then the probability of the event $\{\ln\sum_{i=1}^{e^{nr}} V_i\leq na\}$ decays  doubly exponentially, i.e.,
\begin{equation}\label{eqn:prob_leqdouble}
\Pr\left\{\sum_{i=1}^{e^{nr}} V_i\leq e^{na}\right\}\leq \exp\{-e^{na}(e^{n\epsilon}-n\epsilon-1)\}.
\end{equation}
\end{enumerate}
\end{proposition}
\begin{IEEEproof}[Proof of Proposition \ref{prop:LLN}]
  Part 1  {follows from a clipped version of  Markov's inequality}. See the derivation of~\cite[Eqn.~(41)]{somekh2011exact}. Parts 2) and 3)  {follows by applying the Chernoff bound}. See~\cite[Appendix B]{merhav2009relations}.
\end{IEEEproof}
Base on Fact \ref{fact} and Proposition \ref{prop:LLN}, we can derive the following lemma which is essential in handling the statistical dependence between $F_1$ and $F_4$. Note that, by definition, these random variables share the same ``cloud centers''.
\begin{lemma}\label{lem:dependency}
Given $Y^n=y^n$, $\calC'_U=c'_U$ and $T\geq 0$, for $n$ sufficiently large, we have
\begin{align}\label{eqn:inq_dependency}
\Pr\big\{F_1\!\leq\! e^{-nT}\cdot F_4\,\big|\,Y^n\!=\!  y^n,\calC'_U= c'_U\big\}\le e^{-n\cdot  R_1/4}.
\end{align}
\end{lemma}
\begin{IEEEproof}
Let $\calC'_X:=\{X^n(m_1,m_2):m_1\in \calM_1,m_2\in \calM_2\setminus \{1\}\}$, and define
\begin{equation}\label{def:f_Q}
f(Q_{UXY}):= -\mathbb{E}_{Q_{UXY}}[\ln W_{\calY}(Y|X)].
\end{equation}
Recall the definitions of $\Lambda(Q_{UY},c'_U)$ and $N_i(Q_{UXY})$ (see \eqref{def:U_num} and \eqref{def:XU_num}) and let $Q_{UXY}=Q_{U|Y}Q_{X|UY}\hatQ_Y$ and $Q'_{UXY}=Q'_{U|Y}Q'_{X|UY}\hatQ_Y$, we have 
\begin{align}
&\Pr\{F_1\leq e^{-nT}F_4|Y^n=y^n,\calC'_U=c'_U\}\nn\\
&=\sum_{c'_X}\Pr\{\calC'_X=c'_X|Y^n=y^n,\calC'_U=c'_U\}\nn\\*
&\quad\times\mathbbm{1}\Bigg\{\sum_{Q_{U|Y}}\sum_{Q_{X|UY}}\sum_{i=2}^{e^{nR_1}}N_i(Q_{UXY})e^{-nf(Q_{UXY})}\nn\\
&\quad\quad \leq e^{-nT}\sum_{Q'_{U|Y}}\sum_{Q'_{X|UY}}N_1(Q'_{UXY})e^{-nf(Q'_{UXY})}\Bigg\}.\label{eqn:rewrite_dependency}
\end{align}
Define the   sets 
\begin{align}
  &\calQ_0(y^n,c'_U)  :=\{Q_{U|Y}:\Lambda(Q_{U|Y}\hatQ_{Y},c'_U)\ge 1\},\label{def:Q_0}
  \end{align}
  and\begin{align}
 &\calQ_1(Q_{U|Y},y^n,c'_U,c'_X) :=\{Q_{X|UY}:\exists\, i\in \calM_1,\nn\\
 &\qquad\mbox{s.t.}\;\;\; N_i(Q_{X|UY}Q_{U|Y}\hatQ_{Y})\ge 1\}.
\end{align} 
We have the chain of inequalities~\eqref{eqn:pfdependency_t1}--\eqref{eqn:pfdependency_t3} on the top of the next page, 
\begin{figure*}
\begin{align}
&\mathbbm{1}\left\{\sum_{Q_{U|Y}}\sum_{Q_{X|UY}}\sum_{i=2}^{e^{nR_1}}N_i(Q_{UXY})e^{-nf(Q_{UXY})}\leq e^{-nT}\sum_{Q'_{U|Y}}\sum_{Q'_{X|UY}}N_1(Q'_{UXY})e^{-nf(Q'_{UXY})}\right\}\nn\\
&=\mathbbm{1}\Bigg\{\sum_{Q_{U|Y}\in\calQ_0(y^n,c'_U)}\sum_{Q_{X|UY}\in\calQ_1(Q_{U|Y},y^n,c'_U,c'_X)}\sum_{i=2}^{e^{nR_1}}N_i(Q_{UXY})e^{-nf(Q_{UXY})}\nn\\
&\quad\quad \leq e^{-nT}\sum_{Q'_{U|Y}\in\calQ_0(y^n,c'_U)}\sum_{Q'_{X|UY}\in\calQ_1(Q'_{U|Y},y^n,c'_U,c'_X)}N_1(Q'_{UXY})e^{-nf(Q'_{UXY})}\Bigg\}\label{eqn:pfdependency_t1}\\
&\leq \sum_{Q_{U|Y}\in\calQ_0(y^n,c'_U)}\sum_{Q_{X|UY}\in\calQ_1(Q_{U|Y},y^n,c'_U,c'_X)}\mathbbm{1}\Bigg\{\sum_{i=2}^{e^{nR_1}}N_i(Q_{UXY})e^{-nf(Q_{UXY})} \leq e^{-nT}N_1(Q_{UXY})e^{-nf(Q_{UXY})}\Bigg\}\label{eqn:pfdependency_t2}\\
&=\sum_{Q_{U|Y}\in\calQ_0(y^n,c'_U)}\sum_{Q_{X|UY}\in\calQ_1(Q_{U|Y},y^n,c'_U,c'_X)}\mathbbm{1}\left\{\sum_{i=2}^{e^{nR_1}}N_i(Q_{UXY})\leq e^{-nT}N_1(Q_{UXY})\right\}\\
&=\sum_{Q_{U|Y}}\sum_{Q_{X|UY}}\mathbbm{1}\left\{\sum_{i=2}^{e^{nR_1}}N_i(Q_{UXY})\leq e^{-nT}N_1(Q_{UXY})\right\}\mathbbm{1}\{\Lambda(Q_{U|Y}\hatQ_{Y},c'_U)\geq 1\}\mathbbm{1}\{N_1(Q_{UXY})\geq 1\}\label{eqn:pfdependency_t3}
\end{align} \hrulefill
\end{figure*}
where \eqref{eqn:pfdependency_t1} is from the fact that $\Lambda(Q_{U|Y}\hatQ_{Y},c'_U)$ and $N_i(Q_{UXY})$, $i\in\calM_1$, are non-negative integers, \eqref{eqn:pfdependency_t2} is due to the fact that 
$
   \sum_{i=1}^{k} a_i \leq \sum_{i=1}^{k} b_i $ implies that there exists  an $i\in \{1,\ldots, k\}$ such that $a_i\leq b_i$,
and the last indicator function $\mathbbm{1}\{N_1(Q_{UXY})\geq 1\}$ in \eqref{eqn:pfdependency_t3} is present because when $N_1(Q_{UXY})=0$, we have $\sum_{i=2}^{e^{nR_1}}N_i(Q_{UXY})\geq 1$ since $Q_{X|UY}\in\calQ_1(Q_{U|Y},y^n,c'_U,c'_X)$ and $Q_{U|Y}\in\calQ_0(y^n,c'_U)$.\par
Therefore, combining \eqref{eqn:rewrite_dependency} and \eqref{eqn:pfdependency_t3}, we have:
\begin{align}
&\Pr\big\{F_1\leq e^{-nT}F_4\,\big|\, Y^n=y^n,\calC'_U=c'_U\big\}\nn\\
&\leq \sum_{Q_{U|Y}}\sum_{Q_{X|UY}}\!\mathbb{E}_{C'_X}\bigg[\mathbbm{1}\bigg\{\sum_{i=2}^{e^{nR_1}}N_i(Q_{UXY})\!\leq\! e^{-nT}N_1(Q_{UXY})\bigg\}\nn\\
&\quad\quad \times\mathbbm{1}\{\Lambda(Q_{U|Y}\hatQ_{Y},c'_U)\geq 1\}\mathbbm{1}\{N_1(Q_{UXY})\geq 1\}\,\nn\\*
&\qquad\qquad \bigg|\,Y^n=y^n,\calC'_U=c'_U\bigg]\label{eqn:dependency_upperbound}
\end{align}
Moreover, let $A_n\triangleq \sum_{i=2}^{e^{nR_1}}N_i(Q_{UXY})$. From Part 2 of Fact~\ref{fact}, we know that $A_n$ is a binomial random variable with $\Lambda(Q_{U|Y}\hatQ_{Y},c'_U)(e^{nR_1}-1)$ trials and "success" probability where the corresponding exponent is $(\beta(Q_{UXY})+R_1)$.

There are two cases for the exponent of the expectation of $A_n$, i) $\liminf_{n\to \infty}\frac{1}{n}\ln \mathbb{E}[A_n]>0$, and ii) $\liminf_{n\to \infty}\frac{1}{n}\ln\mathbb{E}[A_n]\leq 0$.\par
For the first case, we know that  for sufficiently large $n$,  
\begin{equation}
B_n:=\frac{1}{n}\ln \Lambda(Q_{U|Y}\hatQ_{Y},c'_U)-\beta(Q_{UXY},R_1)>0
\end{equation}
uniformly.
Then using Parts 2 and 3 of Proposition \ref{prop:LLN}, for any sufficiently small $\epsilon\in (0,B_n)$, we have 
\begin{align}
&\Pr\left\{\frac{1}{n}\ln A_n \leq B_n-\epsilon \bigcup \frac{1}{n}\ln A_n \geq B_n+\epsilon\right\}\nn\\*
&\qquad\leq 2\exp\big\{-e^{n (B_n-\epsilon)}\big\}\label{eqn:A_concentrate}.
\end{align}
In other words, $A_n$ concentrates doubly exponentially fast around its expectation $\mathbb{E}[A_n]$.\par
Therefore, using a similar derivation as in \cite[Eqns.\ (36)--(39)]{merhav2014exact}, for any sufficiently small $\epsilon>0$, we have the chain of inequalities~\eqref{eqn:TIT-t1}--\eqref{eqn:dependency_case1} on the top of the next page,
\begin{figure*}
\begin{align}
&\Pr\left\{\sum_{i=2}^{e^{nR_1}}N_i(Q_{UXY})\leq e^{-nT}N_1(Q_{UXY})\,\bigg|\,Y^n=y^n,\calC'_U=c'_U\right\}\nn\\
&\leq \sum_{j=0}^{(R_1+R_2)/\epsilon} \Pr\left\{j\epsilon\leq \frac{1}{n}\ln A_n\leq (j+1)\epsilon\right\} \Pr\left\{e^{nj\epsilon}\leq e^{-nT}N_1(Q_{UXY})\,\big|\,Y^n=y^n,\calC'_U=c'_U\right\}\label{eqn:TIT-t1}\\
&\doteq \Pr\left\{\mathbb{E}\left[\sum_{i=2}^{e^{nR_1}}N_i(Q_{UXY})\right]\leq e^{-nT}N_1(Q_{UXY})\,\bigg|\,Y^n=y^n,\calC'_U=c'_U\right\}\label{eqn:dependency_case1t1}\\
&\doteq\Pr\left\{N_1(Q_{UXY})\geq e^{nT}e^{nR_1}\mathbb{E}[N_1(Q_{UXY})]\,\Big|\,Y^n=y^n,\calC'_U=c'_U\right\}\label{eqn:dependency_case1t2}\\
&\leq e^{-n(T+R_1)},\label{eqn:dependency_case1}
\end{align}
\hrulefill
\end{figure*}
where~\eqref{eqn:dependency_case1t1} is due to  \eqref{eqn:A_concentrate} and the fact that $\epsilon$ can be made arbitrarily small,~\eqref{eqn:dependency_case1t2} is due to the fact that $N_i(Q_{UXY})$, $i\in\calM_1$ are \iid (see Part 2 of Fact \ref{fact}) and \eqref{eqn:dependency_case1} is due to Markov's inequality.\par
For the second case in which $\mathbb{E}[A_n]$ is not exponentially large, 
we also have that 
\begin{equation}
\mathbb{E}\bigg[\sum_{i=2}^{e^{nR_1}}N_i(Q_{UXY})\bigg]=(e^{nR_1}-1)\mathbb{E}\big[N_1(Q_{UXY})\big].
\end{equation}
Thus, we have
\begin{equation}
\liminf_{n\to \infty}-\frac{1}{n}\ln \mathbb{E}\big[N_1(Q_{UXY})\big] \ge R_1.\label{eqn:liminf}
\end{equation}
Furthermore, for sufficiently large $n$, by using~\eqref{eqn:liminf} and Markov's inequality, we have 
\begin{align}\label{eqn:dependency_case2}
\Pr\big\{N_1(Q_{UXY})\geq 1\,\big|\,Y^n=y^n,\calC'_U=c'_U\big\}\leq e^{-nR_1/2}
\end{align}
Therefore, combining \eqref{eqn:dependency_upperbound}, \eqref{eqn:dependency_case1} and  \eqref{eqn:dependency_case2}, for sufficiently large $n$, we have
\begin{equation}
\Pr\big\{F_1\leq e^{-nT}F_4\,\big|\,Y^n=y^n,\calC'_U=c'_U\big\}\leq e^{-nR_1/4}.
\end{equation}
This concludes the proof of Lemma~\ref{lem:dependency}. 
\end{IEEEproof}
Now, we use Lemma \ref{lem:dependency}  to prove the following lemma which {eliminates $F_4$ from the probability of interest, removes} the dependence between $F_1$ and $F_4$, and also simplifies the calculation of $\Pr\{\calE_1\}$ (see \eqref{eqn:compact_E1}). 

\begin{lemma}\label{lem:remove_dependency}
For given $(U^n(1),X^n(1,1),Y^n)=(u^n,x^n,y^n)$, $\calC'_U=c'_U$ and $T\geq 0$, we have
\begin{align}
&\Pr\left\{{F_1+F_2}>\max\{f_3,F_4\}\cdot e^{-nT}\right\} \nonumber\\*
&\doteq  \max\left\{\Pr\{F_1>f_3\cdot e^{-nT}\},\Pr\{F_2>f_3\cdot e^{-nT}\}\right\}.
\end{align}
\end{lemma}
\begin{IEEEproof}[Proof of Lemma~\ref{lem:remove_dependency}]
See Appendix~\ref{pf:lem_remove_dependency}.
\end{IEEEproof}

{Now, we continue the proof of Theorem~\ref{TH:E1} by using Lemma~\ref{lem:remove_dependency}. Recall the error probability in \eqref{eqn:compact_E1}. Note that }
\begin{align}
\Pr\{\mathcal{E}_1\}
&\doteq  \Pr\left\{F_1+F_2>\max\{f_3,F_4\}\cdot e^{-nT}\right\}\label{pft3} .
\end{align} 
By using Lemma~\ref{lem:remove_dependency}, we have
\begin{align}
&\Pr\{\mathcal{E}_1\}\nn\\*
&\doteq  \max\left\{\Pr\{F_1>f_3\cdot e^{-nT}\},\Pr\{F_2>f_3\cdot e^{-nT}\}\right\}\label{eqn:simplified_error} . 
\end{align}
Next, we consider the first term in the right-hand-side of~\eqref{eqn:simplified_error}. {Recall the definitions of $f(Q_{UXY})$ and $\calQ_0(y^n,c'_U)$ in the proof of Lemma \ref{lem:dependency} (see \eqref{def:f_Q} and \eqref{def:Q_0}) and} let 
\begin{align}
s:= -\frac{1}{n}\ln (f_3\cdot e^{-nT})=f(\hat{Q}_{UXY})+T.
\end{align}
Now, let $Q_{UXY}=Q_{X|UY}Q_{U|Y}\hat{Q}_{Y}$, we have the chain of exponential equalities~\eqref{eqn:TIT-t2}--\eqref{eqn:F1_calc_x} on the top of the next page, 
\begin{figure*}
\begin{align}
&\mathbb{E}_{\mathcal{C}'_U} \left[\Pr\{F_1>f_3\cdot e^{-nT}\big| \mathcal{C}'_U\} \right]\nonumber\\
&= \mathbb{E}_{\mathcal{C}'_U}\left[\Pr\bigg\{\sum_{Q_{U|Y}\in \mathcal{Q}_0(y^n,\calC'_U)}\sum_{Q_{X|UY}}\sum_{i=2}^{e^{nR_1}}N_i(Q_{UXY})e^{-nf(Q_{UXY})}\geq e^{-ns}\,\bigg|\, \mathcal{C}'_U\bigg\}\right]\label{eqn:TIT-t2}\\
&\doteq \mathbb{E}_{\mathcal{C}'_U}\left[\Pr\bigg\{\max_{Q_{U|Y}\in \mathcal{Q}_0(y^n,\calC'_U)}\max_{Q_{X|UY}}\sum_{i=2}^{e^{nR_1}}N_i(Q_{UXY})e^{-nf(Q_{UXY})}\geq e^{-ns}\,\bigg|\, \mathcal{C}'_U\bigg\}\right]\\
&\doteq \mathbb{E}_{\mathcal{C}'_U}\left[\max_{Q_{U|Y}\in \mathcal{Q}_0(y^n,\calC'_U)}\max_{Q_{X|UY}}\Pr\bigg\{\sum_{i=2}^{e^{nR_1}}N_i(Q_{UXY})e^{-nf(Q_{UXY})}\geq e^{-ns}\,\bigg|\, \mathcal{C}'_U\bigg\}\right]\label{eqn:prob_max}\\
&\doteq \mathbb{E}_{\mathcal{C}'_U}\left[\sum_{Q_{U|Y}\in \mathcal{Q}_0(y^n,\calC'_U)}\max_{Q_{X|UY}}\Pr\bigg\{\sum_{i=2}^{e^{nR_1}}N_i(Q_{UXY})e^{-nf(Q_{UXY})}\geq e^{-ns}\,\bigg|\, \mathcal{C}'_U\bigg\}\right]\\
&=\mathbb{E}_{\mathcal{C}'_U}\left[\sum_{Q_{U|Y}}\mathbbm{1}\{\Lambda(Q_{U|Y}\hatQ_{Y},\calC'_U)\geq 1\}\max_{Q_{X|UY}}\Pr\bigg\{\sum_{i=2}^{e^{nR_1}}N_i(Q_{UXY})e^{-nf(Q_{UXY})}\geq e^{-ns}\,\bigg|\, \mathcal{C}'_U\bigg\}\right]\label{eqn:F1_review}\\
&=\sum_{Q_{U|Y}}\mathbb{E}_{\mathcal{C}'_U}\left[\mathbbm{1}\{\Lambda(Q_{U|Y}\hatQ_{Y},\calC'_U)\geq 1\}\max_{Q_{X|UY}}\Pr\bigg\{\sum_{i=2}^{e^{nR_1}}N_i(Q_{UXY})\geq e^{n(f(Q_{UXY})-s)}\,\bigg|\, \mathcal{C}'_U\bigg\}\right]\label{eqn:F1_calc_x}
\end{align}
\hrulefill
\end{figure*}
where  the interchange of $\max\{\cdot\}$ and $\Pr\{\cdot\}$ in \eqref{eqn:prob_max} is justified similarly as \cite[Eqn.~(37)]{somekh2011exact} and~\cite[Eqns.~(15)--(20)]{merhav2014exact}.\par
Using Part 2 of Fact \ref{fact}, for a given $\mathcal{C}_U'=c_U'$, we evaluate the inner probability 
in \eqref{eqn:F1_calc_x} as follows. 
\begin{enumerate}
\item The case $f(Q_{UXY})-s\leq 0$. Note that $N_i(Q_{UXY})$, $i\in \calM_1\setminus \{1\}$, are non-negative integers. Using Part 1 of Proposition~\ref{prop:LLN}, we have
\begin{align}
&\Pr\bigg\{\sum_{i=2}^{e^{nR_1}}N_i(Q_{UXY})\geq e^{n[f(Q_{UXY})-s]}\,\bigg|\, \mathcal{C}'_U=c'_U\bigg\}\nn\\
&=\Pr\bigg\{\sum_{i=2}^{e^{nR_1}}N_i(Q_{UXY})\geq 1\,\bigg|\, \mathcal{C}'_U=c'_U\bigg\}\\
&\doteq \exp\bigg\{-n\Big|\beta(Q_{UXY})-\frac{1}{n}\ln \Lambda(Q_{U|Y}\hatQ_{Y},c'_U)\Big|_+\bigg\}.
\end{align}
\item The case $f(Q_{UXY})-s > \big|\frac{1}{n}\ln \Lambda(Q_{U|Y}\hatQ_{Y},c'_U)-\beta(Q_{UXY})\big|_+$. Using Part 2 of Proposition \ref{prop:LLN}, for sufficiently large $n$, we have
\begin{align}
&\Pr\bigg\{\sum_{i=2}^{e^{nR_1}}N_i(Q_{UXY})\geq e^{n[f(Q_{UXY})-s]}\,\bigg|\, \mathcal{C}'_U=c'_U\bigg\}\nn\\*
&\qquad\leq \exp\big\{-e^{n[f(Q_{UXY})-s]}\big\}.
\end{align}
This term decays at least doubly exponentially rapidly and hence its exponent is infinity.
\item The case $0<f(Q_{UXY})-s < \big[\frac{1}{n}\ln \Lambda(Q_{U|Y}\hatQ_{Y},c'_U)-\beta(Q_{UXY})\big]$. Using Part 3 of Proposition \ref{prop:LLN}, for sufficiently large $n$, we have
\begin{align}
&\Pr\bigg\{\sum_{i=2}^{e^{nR_1}}N_i(Q_{UXY})\geq e^{n[f(Q_{UXY})-s]}\,\bigg|\, \mathcal{C}'_U=c'_U\bigg\}\nn\\
&=1-\Pr\bigg\{\sum_{i=2}^{e^{nR_1}}N_i(Q_{UXY})< e^{n[f(Q_{UXY})-s]}\,\nn\\*
&\qquad\qquad\qquad\qquad\qquad\qquad\qquad \bigg|\, \mathcal{C}'_U=c'_U\bigg\}\\
&\geq 1-\exp\big\{-e^{n[f(Q_{UXY})-s]}\big\}.
\end{align} 
This term converges to 1 at least doubly exponentially fast and hence its exponent is 0.
\end{enumerate}
In summary, we have~\eqref{eqn:E3_bound_detail}--\eqref{eqn:E3_bound} on the top of the next page, 
\begin{figure*}
\begin{align}
&\Pr\bigg\{\sum_{i=2}^{e^{nR_1}}N_i(Q_{UXY})\geq e^{n(f(Q_{UXY})-s)}\,\bigg|\, \mathcal{C}'_U=c'_U\bigg\}\nn\\
&\doteq \exp\left\{-n\left\{
\begin{aligned} 
&\Big|\beta(Q_{UXY})-\frac{1}{n}\ln \Lambda(Q_{U|Y}\hatQ_{Y},c'_U)\Big|_+  &&\text{if} \quad f(Q_{UXY})-s\leq \Big|\frac{1}{n}\ln \Lambda(Q_{U|Y}\hatQ_{Y},c'_U)-\beta(Q_{UXY})\Big|_+\\
&\infty &&\text{if}\quad f(Q_{UXY})-s > \Big|\frac{1}{n}\ln \Lambda(Q_{U|Y}\hatQ_{Y},c'_U)-\beta(Q_{UXY})\Big|_+
\end{aligned}
\right.\right\}\label{eqn:E3_bound_detail}\\
&=\exp\bigg\{-n E_3\Big(Q_{UXY},-s,R_1+\frac{1}{n}\ln\Lambda(Q_{U|Y},c'_U\Big)\bigg\}\label{eqn:E3_bound}
\end{align} \hrulefill
\end{figure*}
where 
\begin{align}\label{def:E3}
&E_3(Q_{UXY},t,r)\nn\\*
&:=\left\{
\begin{aligned}
&|\beta(Q_{UXY},R_1)+R_1-r|_{+}\\
&\qquad\qquad\qquad \mbox{if}\quad Q_{UXY}\in \mathcal{L}_4(t,r,R_1)\\
&\infty \qquad\qquad\quad \text{else}
\end{aligned}
\right.
\end{align}
and where
\begin{align}\label{def:L4}
&\mathcal{L}_4(t,r,R_1)\nn\\
&:= \left\{Q_{UXY}:t+f(Q_{UXY})\leq |r-\beta(Q_{UXY},R_1)-R_1|_{+} \right\}.
\end{align}
Note that the first clause in \eqref{eqn:E3_bound_detail} comes from cases 1) and 3) above. The second clause in \eqref{eqn:E3_bound_detail} comes from case 2) above. For brevity, define
\begin{align}\label{def:E3_opt}
E^*_3(Q_{UY},t,r):=\min_{Q_{X|UY}}E_3(Q_{X|UY}Q_{UY},t,r).
\end{align}
Therefore, recall that $Q_{UXY}=Q_{X|UY}Q_{U|Y}\hat{Q}_{Y}$, by combining~\eqref{eqn:F1_calc_x}, \eqref{eqn:E3_bound} and \eqref{def:E3_opt}, we have the exponential equalities~\eqref{eqn:TIT-t3}--\eqref{eqn:TIT-t4} on the top of the next page.
\begin{figure*}
\begin{align}
&\mathbb{E}_{\mathcal{C}'_U} \left[\Pr\{F_1>f_3\cdot e^{-nT}\big| \mathcal{C}'_U\} \right]\nonumber\\
&\doteq\sum_{Q_{U|Y}}\mathbb{E}_{\mathcal{C}'_U}\left[\mathbbm{1}\{\Lambda(Q_{U|Y}\hatQ_{Y},\calC'_U)\geq 1\}\max_{Q_{X|UY}} \exp\Big\{-n E_3 \Big(Q_{UXY},-s,R_1+\frac{1}{n}\ln\Lambda(Q_{U|Y}\hatQ_{Y},\calC'_U\Big)\Big\}\right]\label{eqn:TIT-t3}\\
&=\sum_{Q_{U|Y}}\mathbb{E}_{\mathcal{C}'_U}\left[\mathbbm{1}\{\Lambda(Q_{U|Y}\hatQ_{Y},\calC'_U)\geq 1\} \exp\Big\{-n E^*_3\Big(Q_{U|Y}\hatQ_Y,-s,R_1+\frac{1}{n}\ln\Lambda(Q_{U|Y}\hatQ_{Y},\calC'_U\Big)\Big\}\right]\\
&\doteq\max_{Q_{U|Y}}\mathbb{E}_{\mathcal{C}'_U}\left[\mathbbm{1}\{\Lambda(Q_{U|Y}\hatQ_{Y},\calC'_U)\geq 1\} \exp\Big\{-n E^*_3\Big(Q_{U|Y}\hatQ_Y,-s,R_1+\frac{1}{n}\ln\Lambda(Q_{U|Y}\hatQ_{Y},\calC'_U\Big)\Big\}\right]\label{eqn:TIT-t4}
\end{align}
\hrulefill
\end{figure*}

Now we regard $\mathcal{C}'_U$ as a collection of random variables. Consequently, $\Lambda(Q_{U|Y}\hatQ_{Y},\calC'_U)$ is also a random variable. Using a similar derivation as in \cite[Eqns.~(36)--(39)]{merhav2014exact}, for any sufficiently small $\epsilon>0$, we have
\begin{align}
&\mathbb{E}_{\mathcal{C}'_U} \left[\Pr\{F_1>f_3\cdot e^{-nT}\big| \mathcal{C}'_U\} \right]\nonumber\\
&\doteq\max_{Q_{U|Y}}\sum_{1\leq\lambda\leq e^{nR_2}} \Pr\{\Lambda(Q_{U|Y}\hatQ_{Y},\calC'_U)=\lambda\}\nn\\*
&\qquad\times\exp\Big\{-n E^*_3\Big(Q_{U|Y}\hatQ_Y,-s,R_1+\frac{1}{n}\ln\lambda\Big)\Big\}\\
&\geq \max_{Q_{U|Y}}\sum_{0\leq i\leq R_2/\epsilon}\nn\\*
& \qquad\Pr \left\{i\epsilon\leq \frac{1}{n}\ln\Lambda(Q_{U|Y}\hatQ_{Y},\calC'_U)\leq (i+1)\epsilon \right\}\nn\\*
&\qquad\qquad\times\exp\big\{-n E^*_3(Q_{U|Y}\hatQ_Y,-s,R_1+i\epsilon)\big\}\label{eqn:F1_calc_u}
\end{align}
Recalling Part 1 of Fact \ref{fact}, we can evaluate the probability $\Pr\{i\epsilon\leq \frac{1}{n}\ln\Lambda(Q_{U|Y}\hatQ_{Y},\calC'_U)\leq (i+1)\epsilon\}$ as follows. 
\begin{enumerate}
\item Case $\gamma(Q_{U|Y}\hatQ_Y,R_2)<0$: Similarly as before, we see that $\Lambda(Q_{U|Y}\hatQ_{Y},\calC'_U)$ concentrates doubly exponentially around its expectation which is, on the exponential scale, $e^{-n\gamma(Q_{U|Y}\hatQ_Y,R_2)}$. In   other words, we have
\begin{align}
&\Pr\left\{i\epsilon\leq \frac{1}{n}\ln\Lambda(Q_{U|Y}\hatQ_{Y},\calC'_U)\leq (i+1)\epsilon\right\}\nn\\*
&\doteq \mathbbm{1}\{i\epsilon \leq -\gamma(Q_{U|Y}\hatQ_Y,R_2) \leq (i+1)\epsilon\}\label{eqn:Lambad_review}
\end{align}
\item Case $\gamma(Q_{U|Y}\hatQ_Y,R_2)\geq 0$: Similarly as before, we see that, on the one hand, 
\begin{align}
&\Pr\left\{\frac{1}{n}\ln\Lambda(Q_{U|Y}\hatQ_{Y},\calC'_U)\geq 0\right\}\nn\\*
&\qquad\doteq \exp\{-n\gamma(Q_{U|Y}\hatQ_Y,R_2)\};
\end{align}
on the other hand, 
\begin{equation}
\Pr\left\{\frac{1}{n}\ln\Lambda(Q_{U|Y}\hatQ_{Y},\calC'_U)\geq \epsilon\right\}\dotleq \exp\{-e^{n\epsilon}\}.
\end{equation}
Therefore, we have
\begin{align}
&\Pr\left\{i\epsilon\leq \frac{1}{n}\ln\Lambda(Q_{U|Y}\hatQ_{Y},\calC'_U)\leq (i+1)\epsilon\right\}\nn\\*
&\qquad\doteq \mathbbm{1}\{i=0\}\exp\{-n\gamma(Q_{U|Y}\hatQ_Y,R_2)\}
\end{align}
\end{enumerate}

In summary, we have
\begin{align}
&\Pr\left\{i\epsilon\leq \frac{1}{n}\ln\Lambda(Q_{U|Y}\hatQ_{Y},\calC'_U)\leq (i+1)\epsilon \right\}\nn\\
&\qquad\doteq \mathbbm{1}\{i\epsilon \leq |-\gamma(Q_{U|Y}\hatQ_Y,R_2)|_+ \leq (i+1)\epsilon\}\nn\\*
&\qquad\qquad\times\exp\{-n|\gamma(Q_{U|Y}\hatQ_Y,R_2)|_+\}\label{eqn:Lambda_evaluate}
\end{align}
{Therefore, by recalling the definitions of $\Phi(Q_{UXY},R_1,R_2)$ and $\mathcal{L}_1(\hat{Q}_{XY},R_1,R_2,T)$ (see \eqref{def:Phi} and \eqref{def:L1}),} and combining \eqref{eqn:F1_calc_u} and \eqref{eqn:Lambda_evaluate}, we have the exponential equalities~\eqref{eqn:F1_temp}--\eqref{eqn:F1_Phi} on the top of the next page,\footnote{We use the notation  $\doteq$ (i.e., equality to first-order in the exponent)  in~\eqref{eqn:F1_temp} since the other direction of the inequality in~\eqref{eqn:F1_calc_u} can be derived by  replacing $i\epsilon$ with $(i+1)\epsilon$ in the function $E_3^*(\cdot)$.   See \cite[Eqns.\ (36)--(39)]{merhav2014exact} for another instance of this calculation.} 
\begin{figure*}
\begin{align}
&\mathbb{E}_{\mathcal{C}'_U} \left[\Pr\{F_1>f_3\cdot e^{-nT}|\calC'_U\} \right]\nonumber\\
&\doteq \max_{Q_{U|Y}}\bigg[\exp\{-nE^*_3(Q_{U|Y}\hatQ_Y,-s,R_1+|-\gamma(Q_{U|Y}\hatQ_Y,R_2)|_{+})\}\cdot \exp\{-n|\gamma(Q_{U|Y}\hatQ_Y,R_2)|_{+}\}\bigg]\label{eqn:F1_temp}\\
&\doteq \exp\bigg\{-n\min_{Q_{U|Y}}\Big[E^*_3(Q_{U|Y}\hatQ_Y,-s,R_1+|-\gamma(Q_{U|Y}\hatQ_Y,R_2)|_{+})+|\gamma(Q_{U|Y}\hatQ_Y,R_2)|_{+}\Big]\bigg\}\\
&= \exp\bigg\{-n\min_{Q_{UX|Y}}\Big[E_3(Q_{UX|Y}\hat{Q}_Y,-s,R_1+|-\gamma(Q_{U|Y}\hatQ_Y,R_2)|_{+})+|\gamma(Q_{U|Y}\hatQ_Y,R_2)|_{+}\Big]\bigg\}\\
&= \exp\bigg\{-n\min_{Q_{UX|Y}\in \mathcal{L}_1(\hat{Q}_{XY},R_1,R_2,T)} \Phi(Q_{UX|Y}\hat{Q}_{Y},R_1,R_2)\bigg\}\label{eqn:F1_Phi},
\end{align}
\hrulefill
\end{figure*}
where~\eqref{eqn:F1_temp} is due to~\eqref{eqn:Lambda_evaluate} and the fact that $\epsilon$ can be made arbitrarily small and~\eqref{eqn:F1_Phi} is due to the fact that
\begin{align}
&|\beta(Q_{UXY},R_1)-|-\gamma(Q_{UXY},R_2)|_{+}|_{+}+|\gamma(Q_{UXY},R_2)|_{+}\nn\\*
&=\left\{
\begin{aligned}
&\beta(Q_{UXY},R_1)+\gamma(Q_{UXY},R_2)\\
&\qquad \mbox{if}\quad \gamma(Q_{UXY},R_2)\geq 0,\beta(Q_{UXY},R_1)\geq 0\\
&\gamma(Q_{UXY},R_2)\\
&\qquad \mbox{if}\quad \gamma(Q_{UXY},R_2)\geq 0,\beta(Q_{UXY},R_1)\leq 0\\
&\beta(Q_{UXY},R_1)+\gamma(Q_{UXY},R_2)\\
&\qquad \mbox{if}\quad \gamma(Q_{UXY},R_2)\leq 0,\\
&\qquad\qquad \mbox{and}\quad \beta(Q_{UXY},R_1)+\gamma(Q_{UXY},R_2)\geq 0\\
&0\quad\ \, \mbox{if}\quad \gamma(Q_{UXY},R_2)\leq 0,\\
&\qquad\qquad \mbox{and}\quad \beta(Q_{UXY},R_1)+\gamma(Q_{UXY},R_2)\leq 0\\
\end{aligned}
\right.\\*
&=\Phi(Q_{UXY},R_1,R_2).\label{eqn:Phi_review}
\end{align}

After averaging over $(U^n(1),X^n(1,1),Y^n)$, we have  
\begin{align}\label{th1pf3}
&\lim_{n\to\infty}-\frac{1}{n}\ln\mathbb{E}_{(U^n(1),X^n(1,1),Y^n)}\nn\\*
&\qquad\qquad\Big[\mathbb{E}_{\mathcal{C}'_U}[\Pr\{F_1>f_3\cdot e^{-nT}\big|\, \mathcal{C}'_U\}]\Big]\nonumber\\*
&=\min_{\hat{Q}_{UXY}}\Big[D(\hat{Q}_{UXY}\|P_{UXY}) \nn\\*
&\quad +\min_{Q_{UX|Y}\in \mathcal{L}_1(\hat{Q}_{XY},R_1,R_2,T)}\Phi(Q_{UX|Y}\hat{Q}_Y,R_1,R_2)\Big].
\end{align}
For the remaining term  $\mathbb{E}_{(U^n(1),X^n(1,1),Y^n)} [\mathbb{E}_{\mathcal{C}'_U}[\Pr\{F_2>f_3\cdot e^{-nT}\big|\, \mathcal{C}'_U\}] ]$ (second term in~\eqref{eqn:simplified_error}), the proof is similar to the proof for~\eqref{th1pf3}, therefore we will only provide an outline. {Recalling the definition of $N(Q_{UXY})$ in~\eqref{def:X_num} and Part 3 of Fact \ref{fact},} we have
\begin{align}
&\mathbb{E}_{\mathcal{C}'_U}\big[\Pr\{F_2>f_3\cdot e^{-nT}\big| \mathcal{C}'_U\}\big] \nonumber\\
&= \Pr\bigg\{\sum_{Q_{X|UY}}N(Q_{X|UY}\hat{Q}_{UY})e^{-nf(Q_{X|UY}\hat{Q}_{UY})}\geq e^{-ns}\bigg\}\label{eqn:F2_review}\\
&\doteq \exp\Big\{-n\min_{Q_{X|UY}}E_3(Q_{X|UY}\hat{Q}_{UY},-s,R_1)\Big\},
\end{align}
After averaging over $(U^n(1),X^n(1,1),Y^n)$, we have 
\begin{align} 
&\lim_{n\to\infty}-\frac{1}{n}\ln\mathbb{E}_{(U^n(1),X^n(1,1),Y^n)}\nn\\*
&\qquad\qquad\qquad\Big[\mathbb{E}_{\mathcal{C}'_U}[\Pr\{F_2>f_3\cdot e^{-nT}\big|\, \mathcal{C}'_U\}]\Big]\nn\\*
&=\Psi_{\mathrm{b}}.
\end{align} 
Then, due to (\ref{eqn:simplified_error}), we have $E_1^{\mathrm{t}}=\min\{\Psi_{\mathrm{a}},\Psi_{\mathrm{b}}\}$.\par
For the total error probability of the message pair, according to the optimal decoding region (\ref{def:DRmp}), we obtain
\begin{align}\label{eqn:errorY}
\mathbb{E}_{\mathcal{C}} [e_Y^{\mathrm{t}}(1,1)]=\mathbb{E}_{\mathcal{C}} \left[\Pr\{F'_1+F_2>f_3\cdot e^{-nT}\big|{\mathcal{C}}\}\right],
\end{align}
where
\begin{align}
F'_1&:= \sum_{m'_1\in \mathcal{M}_1}\sum_{m'_2\in \mathcal{M}_2\setminus\{1\}} W_{\mathcal{Y}}^n(Y^n|X^n(m'_1,m'_2))\\
&=F_1+F_4.\label{def:F_11}
\end{align}
As the difference between $F_1$ and $F'_1$ is only in the number of $m'_1$ (the difference is  exactly one and the rates are asymptotically equal), the exponents of $\mathbb{E}_{\mathcal{C}'_U}\big[\Pr\{F_1>f_3\cdot e^{-nT}\big| \mathcal{C}'_U\}\big]$  and $\mathbb{E}_{\mathcal{C}'_U}\big[\Pr\{F'_1>f_3\cdot e^{-nT}\big| \mathcal{C}'_U\}\big]$   are identical. Therefore, we have $E_Y^{\mathrm{t}}=E_1^{\mathrm{t}}$.\par

Now we explain why $E_1^{\mathrm{u}}=E_1^{\mathrm{t}}+T$ and $ E_Y^{\mathrm{u}}=E_Y^{\mathrm{t}}+T$. In~\cite[Lemma~1]{huleihel2016erasure}, it was shown that, for discrete memoryless channels $W:\calX\to\calY$, the undetected error exponent $E^{\rmu}$ is equal to the sum of the total error exponent $E^{\rmt}$ and the threshold $T$. The main argument is based on the fact that the optimal decoding region 
\begin{equation}
\calD^*_{m}:=\left\{y^n:\frac{W^n(y^n|x^n(m))}{\sum_{m'\neq m}W^n(y^n|x^n(m'))}\geq e^{nT}\right\}
\end{equation} 
minimizes the following function
\begin{equation}
\Gamma(\calC,\calD)=e^{\rmu}+e^{-nT}e^{\rmt}
\end{equation}
for a given codebook $\calC$ and a given threshold $T$, where $e^{\rmu}$ and $e^{\rmt}$ are the average total and undetected error probabilities, respectively. Moreover, the proof of ~\cite[Lemma~1]{huleihel2016erasure} does not depend on the structure of the codebook and the closed-form expressions of the exponents $E^{\rmt}$ and $E^{\rmu}$. Therefore, we can use the same idea to show that $E_1^{\mathrm{u}}=E_1^{\mathrm{t}}+T$ and $ E_Y^{\mathrm{u}}=E_Y^{\mathrm{t}}+T$ since the optimal decoding regions $\calD^*_{m_1}$ and $\calD^*_{m_1m_2}$ defined in \eqref{def:DRj} and \eqref{def:DRmp} also minimize     $\Gamma(\calC,\calD)$ for the ABC. \par

 For   constant composition random codes, since $(U^n(m_2),X^n(m_1,m_2))\in \calT_{P_{UX}}$ for all  $ (m_1,m_2)\in\calM_1\times \calM_2$, all joint types $Q_{UXY}$ (resp.\ $Q_{UY}$) must satisfy  the condition that their marginal distributions $Q_{UX}$ (resp.\ $Q_{U}$) are $P_{UX}$ (resp.\ $P_{U}$). Therefore, the results can be proved similarly to the case for  \iid random codes, except that all types $Q_{UXY}$ (resp.\ $Q_{UY}$) must additionally satisfy the condition that their marginal distributions $Q_{UX}$ (resp.\ $Q_{U}$) are $P_{UX}$ (resp.\ $P_{U}$).
\par 
 
This concludes the proof of Theorem~\ref{TH:E1}. \end{IEEEproof}
\section{Proof of Theorem~\ref{TH:E2}}\label{sec:pfTHE2}
\begin{IEEEproof}[Proof of Theorem~\ref{TH:E2}]
Firstly, we consider   i.i.d.\ random codes. Assume the true transmitted message pair is $(m_1,m_2)=(1,1)$. Define the (total) error event $\mathcal{E}_2$ as
\begin{align}
\mathcal{E}_2:=\bigg\{\sum_{m'_2\neq 1}\Pr(Y^n|\mathcal{C}_2(m'_2))  > \Pr(Y^n|\mathcal{C}_2(1))e^{-nT}\bigg\}.
\end{align}
The average total error probability for message $m_2=1$  associated to the decoding region $\calD^*_{m_2}$ in (\ref{def:DRj}) is given by
\begin{align}
&\mathbb{E}_{\mathcal{C}}[e_2^{\mathrm{t}}(1,1)]\nn\\*
&=\mathbb{E}_{(U^n(1),X^n(1,1),Y^n)} \left[\Pr\{\mathcal{E}_2|(U^n(1),X^n(1,1),Y^n)\}\right].\label{eqn:E2_calc}
\end{align}
Recall the definitions of $F'_1$ and $F_2$ (see \eqref{def:F_11} and \eqref{def:F2}). Similarly, for given $(U^n(1),X^n(1,1),Y^n)=(u^n,x^n,y^n)$ with joint  type $\hat{Q}_{UXY}$, we have 
\begin{align}
\Pr\{\mathcal{E}_2\} &=\Pr\{F'_1>(f_3+F_2)\cdot e^{-nT}\}\label{eqn:E2_compact}.
\end{align} 
For a given sub-codebook $\calC'_2(1):=\{X^n(m_1,1):m_1\in\calM_1\setminus \{1\}\}$, let
\begin{equation}\label{def:small_k}
k:=\frac{1}{n}\ln(f_3+F_2)=\frac{1}{n}\ln\sum_{m_1\in\calM_1}W^n_{\calY}(y^n|x^n(m_1,1))
\end{equation}
and so, the right-hand-side of the inequality inside the probability of \eqref{eqn:E2_compact} is constant. Similarly to the calculation of  $\mathbb{E}_{\mathcal{C}'_U} \left[\Pr\{F_1>f_3\cdot e^{-nT}|\calC'_U\} \right]$ in~\eqref{eqn:F1_Phi}, we obtain 
\begin{align}
&\Pr \left\{F'_1>e^{n(k-T)}\right\}\nn\\*
&\qquad\doteq \exp\{-nE_4(\hatQ_Y,k-T,R_1,R_2)\}\label{eqn:F_11_larger}
\end{align}
where
\begin{align}
&E_4(\hatQ_Y,t,R_1,R_2)\\
&:=\min_{Q_{UX|Y}}\bigg[E_3\left(Q_{UX|Y}\hatQ_Y,t,R_1+|-\gamma(Q_{UX|Y}\hatQ_Y,R_2)|_{+} \right)\nn\\*
&\qquad\qquad +|\gamma(Q_{UX|Y}\hatQ_Y,R_2)|_{+}\bigg]\\
&=\min_{Q_{UX|Y}\in \calL_5(\hatQ_Y,t,R_1,R_2)} \Psi(Q_{UX|Y}\hatQ_Y,R_1,R_2)\label{def:E4}
\end{align}
and where
\begin{align}
&\calL_5(\hatQ_Y,t,R_1,R_2):=\Big\{Q_{UX|Y}:\nn\\*
&\quad\mathbb{E}_{Q_{UX|Y}\hatQ_Y}\ln\frac{1}{W_{\mathcal{Y}}}+t \leq \Delta(Q_{UX|Y}\hatQ_Y,R_1,R_2)\Big\}.\label{def:L5}
\end{align}
Next, we consider the scenario in which the sub-codebook $\mathcal{C}'_2(1)$ is random. Consequently,  
\begin{equation}\label{def:K}
K=\frac{1}{n}\ln(f_3+F_2)
\end{equation}
is also random. Using a similar derivation as in \cite[Eqns.~(36)--(39)]{merhav2014exact}, for any sufficiently small $\epsilon>0$, we have
\begin{align}
&\Pr\left\{F'_1>(f_3+F_2)\cdot e^{-nT}\right\}\nonumber\\
&\doteq\sum_{k}\Pr \left\{K=k \right\}\cdot\exp\{-nE_4(\hatQ_Y,k-T,R_1,R_2)\}\\
&\dotleq \sum_{i}\Pr \left\{i\epsilon\leq K< (i+1)\epsilon\right\}\nn\\*
&\qquad\times\exp\{-nE_4(\hatQ_Y,i\epsilon-T,R_1,R_2)\},\label{eqn:F1_sum}
\end{align}
where $i$ in the last inequality ranges from $-f(\hatQ_{UXY})/\epsilon$ to $R_2/\epsilon$.\par
Recall the definition of $N(Q_{UXY})$ (see \eqref{def:X_num}), and let $Q_{UXY}=Q_{X|UY}\hatQ_{UY}$, we have 
\begin{align}
e^{nk}=e^{-nf(\hatQ_{UXY})}+\sum_{Q_{X|UY}}N(Q_{UXY})e^{-nf(Q_{UXY})}.\label{eqn:k_sum}
\end{align}
Note that the first term in the right side of \eqref{eqn:k_sum} is fixed. For the second term, we now evaluate the following probability 
\begin{align}
\Pr\Big\{e^{nt}\leq \sum_{Q_{X|UY}}N(Q_{UXY})e^{-nf(Q_{UXY})}\leq e^{n(t+\epsilon)}\Big\}.\label{eqn:K1_expansion}
\end{align}
On the one hand, we have (similarly as before) 
\begin{align}
&\Pr\Big\{\sum_{Q_{X|UY}}N(Q_{UXY})e^{-nf(Q_{UXY})}\geq e^{nt}\Big\}\nn\\*
&\doteq \exp\{-nE^*_3(\hatQ_{UY},t,R_1)\}
\end{align}
On the other hand, by using a similar derivation as in \cite[Eqns.~(30)--(34)]{merhav2014exact} and~\cite[pp.~5081]{averbuch2018exact}, we can derive the exponent of the probability of that $\sum_{Q_{X|UY}} N(Q_{UXY}) e^{-nf(Q_{UXY})}$ is upper bounded by $e^{n(t+\epsilon)}$ in the following steps. Firstly, we have
\begin{align}
&\Pr\Big\{\sum_{Q_{X|UY}}N(Q_{UXY})e^{-nf(Q_{UXY})}\leq e^{n(t+\epsilon)}\Big\}\nn\\
&\doteq \Pr\Big\{\max_{Q_{X|UY}}N(Q_{UXY})e^{-nf(Q_{UXY})}\leq e^{n(t+\epsilon)}\Big\}\label{eqn:prob_NC_reply}\\
&\doteq \Pr\bigg\{\! \bigcap_{Q_{X|UY}}\! \Big\{N(Q_{UXY})\! \leq \! \exp\{n[t\! +\! \epsilon\! +\!  f(Q_{UXY})]\}\Big\}\bigg\}\label{eqn:prob_NC_larger}.
\end{align}
Recall Part 3 of Fact \ref{fact}, there are two cases for the probability of the events $\{N(Q_{UXY})\leq \exp\{n[t+\epsilon+f(Q_{UXY})]\}\}$:
\begin{enumerate}
\item Case $\beta(Q_{UXY},R_1)<0$ and $[t+\epsilon+f(Q_{UXY})]<-\beta(Q_{UXY},R_1)$. From Part 3 of Proposition \ref{prop:LLN}, we see that
\begin{equation}
\Pr\big\{N(Q_{UXY})\leq \exp\{n[t+\epsilon+f(Q_{UXY})]\}\big\}\doteq 0
\end{equation}
\item Case $\beta(Q_{UXY},R_1)> 0$ or $[t+\epsilon+f(Q_{UXY})]\geq -\beta(Q_{UXY},R_1)$. Similarly as before, for  sufficiently large $n$, we have
\begin{align}
&\Pr\big\{N(Q_{UXY})\leq \exp\{n[t+\epsilon+f(Q_{UXY})]\}\big\}\nn\\
&=1-\Pr\big\{N(Q_{UXY})> \exp\{n[t+\epsilon+f(Q_{UXY})]\}\big\}\\
&\geq 1-\exp\{-n|\beta(Q_{UXY},R_1)|\}\,{\to 1}\label{eqn:rvtemp1}
\end{align}
\end{enumerate}
Therefore, the probability in \eqref{eqn:prob_NC_larger} is on the exponential scale equal to the indicator function which returns $1$ if for every $Q_{X|UY}$, either $\beta(Q_{UXY},R_1)> 0$ or $[t+\epsilon+f(Q_{UXY})]\geq -\beta(Q_{UXY},R_1)$, or equivalently,
\begin{align}
&\Pr\Big\{\sum_{Q_{X|UY}}N(Q_{UXY})\leq e^{n[t+\epsilon+f(Q_{UXY})]}\Big\}\nn\\
&\doteq \mathbbm{1}\left\{\min_{Q_{X|UY}}\Big\{\beta(Q_{UXY},R_1)+\big|t+\epsilon+f(Q_{UXY})\big|_+\Big\}\geq 0\right\}\label{eqn:prob_NC_smaller}
\end{align}
We now find the minimum value of $ t+\epsilon $ for which the value of this indicator function is unity. The condition in the indicator function above is equivalent to 
\begin{equation}
 \min_{Q_{X|UY}}\max_{0\leq a\leq 1}\{\beta(Q_{UXY},R_1)+a[t+\epsilon+f(Q_{UXY})]\}\geq 0
\end{equation} 
or, equivalently:
\begin{align}
&\forall \,Q_{X|UY}\:\exists \, 0\leq a\leq 1:\nn\\*
&\qquad \beta(Q_{UXY},R_1)+a[t+\epsilon+f(Q_{UXY})]\geq 0,
\end{align}
which can also be written as
\begin{align}
&\forall \, Q_{X|UY}\:\exists\, 0\leq a\leq 1: \nn\\*
&\qquad\qquad t+\epsilon\geq -f(Q_{UXY})-\frac{\beta(Q_{UXY},R_1)}{a}.
\end{align}
This is equivalent to
\begin{align}
&t+\epsilon\nn\\*
&\geq \max_{Q_{X|UY}}\min_{0\leq a\leq 1}\left[-f(Q_{UXY})-\frac{\beta(Q_{UXY},R_1)}{a}\right]\\
&=\max_{Q_{X|UY}}\Bigg[-f(Q_{UXY})\nn\\*
&\qquad-\left\{
\begin{aligned}
&\beta(Q_{UXY},R_1)\quad &&\beta(Q_{UXY},R_1)\leq 0\\
&\infty &&\beta(Q_{UXY},R_1)>0
\end{aligned}
\right.
\Bigg]\\
&=-\min_{Q_{X|UY}:\beta(Q_{UXY},R_1)\leq 0} [f(Q_{UXY})+\beta(Q_{UXY},R_1)]\label{eqn:s0_expression}\\
&=s_0(\hatQ_{UY},R_1)\label{eqn:s0}
\end{align}
where the minimum in \eqref{eqn:s0_expression} over an empty set is defined as infinity.\par
Furthermore, we need the following lemma which provides  some useful properties of $s_0(\hatQ_{UY},R_1)$ defined in \eqref{def:s0} (also see \eqref{eqn:s0}) and $E^*_3(\hatQ_{UY},t,R_1)$ defined in \eqref{def:E3_opt}. Using this lemma, we can obtain the exponent of the probability in \eqref{eqn:K1_expansion}.
\begin{lemma}\label{lem:E3_vinish}
\begin{enumerate}
\item $s_0(\hatQ_{UY},R_1)>-\infty$, i.e., the set $\{Q_{X|UY}:\beta(Q_{X|UY}\hatQ_{UY},R_1)\leq 0\}$ is not empty.
\item $E^*_3(\hatQ_{UY},t,R_1)$ vanishes for all $t\leq s_0(\hatQ_{UY},R_1)$.
\item $E^*_3(\hatQ_{UY},t,R_1)$ is strictly positive for all $t>s_0(\hatQ_{UY},R_1)$.
\end{enumerate}
\end{lemma}
\begin{IEEEproof}[Proof of Lemma~\ref{lem:E3_vinish}]
See Appendix~\ref{pf:lem_E3_vinish}.
\end{IEEEproof}
In summary, we have
\begin{align}
&\Pr\Big\{\sum_{Q_{X|UY}}N(Q_{UXY})e^{-nf(Q_{UXY})}<e^{n [s_0(\hatQ_{UY},R_1)-\epsilon]}\Big\}\nn\\*
&\doteq 0\label{eqn:K1_leq}\\
&\Pr\Big\{\sum_{Q_{X|UY}}N(Q_{UXY})e^{-nf(Q_{UXY})}\geq e^{n[s_0(\hatQ_{UY},R_1)+\epsilon]}\Big\}\nn\\*
&\doteq \exp\{-nE^*_3(\hatQ_{UY},s_0(\hatQ_{UY},R_1)+\epsilon,R_1)\}\label{eqn:K1_geq}
\end{align}
Furthermore, by using Lemma \ref{lem:E3_vinish}, we conclude that 
\begin{align}
&\Pr\Big\{e^{n[s_0(\hatQ_{UY},R_1)-\epsilon]}\leq \sum_{Q_{X|UY}}N(Q_{UXY})e^{-nf(Q_{UXY})}\nn\\*
&\qquad\qquad\qquad\qquad\qquad< e^{n[s_0(\hatQ_{UY},R_1)+\epsilon]}\Big\}\doteq 1\label{eqn:K1_mid}
\end{align}
Therefore, we have
\begin{align}
&\Pr\left\{F'_1>(f_3+F_2)\cdot e^{-nT}\right\}\nn\\
&\dotleq \sum_{i}\Pr\Big\{e^{ni\epsilon}\!\leq\! \sum_{Q_{X|UY}}\!\! N(Q_{UXY})e^{-nf(Q_{UXY})}\!\leq\! e^{n[(i+1)\epsilon]}\Big\}\nn\\
&\quad \times\! \exp\Big\{\!-nE_4\big(\hatQ_Y,\max\{i\epsilon,-f(\hatQ_{UXY})\}\!-\! T,R_1,R_2 \big)\Big\}\label{eqn:E2_E3temp}
\end{align}
where the expression $\max\{i\epsilon,-f(\hatQ_{UXY})\}$ in the argument of $E_4(\hatQ_Y,\cdot,R_1,R_2)$ is due to the fact that
\begin{align}
K&=\frac{1}{n}\ln\bigg[ e^{-nf(\hatQ_{UXY})}+\sum_{Q_{X|UY}}N(Q_{UXY})e^{-nf(Q_{UXY})}\bigg]\\
&\geq \frac{1}{n}\ln\left[ e^{-nf(\hatQ_{UXY})}+e^{ni\epsilon}\right]\\
&\doteq \max\{i\epsilon,-f(\hatQ_{UXY})\}
\end{align}
By using the fact that $\epsilon$ above can be made arbitrarily small, we obtain
\begin{align}
&\Pr\left\{F'_1>(f_3+F_2)\cdot e^{-nT}\right\}\nn\\
&\doteq \exp\big\{-nE_4(\hatQ_Y,\max\{s_0(\hat{Q}_{UY},R_1),-f(\hatQ_{UXY})\}-T,\nn\\*
&\qquad\qquad\qquad\qquad\qquad\qquad\qquad\qquad\quad R_1,R_2)\big\}\label{eqn:E2_E3final}
\end{align}
where \eqref{eqn:E2_E3final} is due to the fact that the dominant contribution to the sum over $i$ is due to the term indexed by $i=s_0(\hatQ_{UY},R_1)/\epsilon$. This, itself, follows from \eqref{eqn:K1_mid} and \eqref{eqn:E2_E3temp} 
 as well as the fact that $t\mapsto E_4(\hatQ_Y,t,R_1,R_2)$, as defined in \eqref{def:E4}, is non-decreasing.\par
Note that $s_0(\hat{Q}_{UY},R_1)$ and $f(\hatQ_{UXY})$ are constant (given $(U^n(1),X^n(1,1),Y^n)=(u^n,x^n,y^n)$). Once again, by using the fact that the function $E_4(\hatQ_Y,t,R_1,R_2)$, as defined in \eqref{def:E4}, is non-decreasing in the parameter $t$, we have
\begin{align}
&E_4(\hatQ_Y,\max\{s_0(\hat{Q}_{UY},R_1),-f(\hatQ_{UXY})\}-T,R_1,R_2)\nn\\
&=\max\big\{E_4(\hatQ_Y,-f(\hatQ_{UXY})-T,R_1,R_2),\nn\\*
&\qquad\qquad\qquad E_4(\hatQ_Y,s_0(\hat{Q}_{UY},R_1)-T,R_1,R_2)\big\}\label{eqn:E2_final_0}\\
&=\max\Bigg\{\min_{ \substack{Q_{UX|Y} \\ \in \mathcal{L}_1(\hat{Q}_{XY},R_1,R_2,T)}} \Phi(Q_{UX|Y}\hat{Q}_{Y},R_1,R_2),\nn\\*
&\qquad\qquad\min_{\substack{Q_{UX|Y}\\\in \mathcal{L}_3(\hat{Q}_{UXY},R_1,R_2,T)} }\Phi(Q_{UX|Y}\hat{Q}_{Y},R_1,R_2)\Bigg\}\label{eqn:E2_final}
\end{align}
%
By combining~\eqref{eqn:E2_calc}, \eqref{eqn:E2_E3final} and~\eqref{eqn:E2_final} and averaging over $(U^n(1),X^n(1,1),Y^n)$, we have 
\begin{align}
&\lim_{n\to\infty}-\frac{1}{n}\ln\mathbb{E}_{\mathcal{C}}[e_2^{\mathrm{t}}(1,1)]\nn\\*
&=\lim_{n\to\infty}-\frac{1}{n}\ln\mathbb{E}_{(U^n(1),X^n(1,1),Y^n)}\nn\\*
&\qquad\qquad\qquad \big[\Pr\{\mathcal{E}_2|(U^n(1),X^n(1,1),Y^n)\}\big]\\
&=\max\{\Psi_{\mathrm{a}},\Psi_{\mathrm{c}}\}.\label{eqn:E2}
\end{align}
Finally,  the equality $E_2^{\mathrm{u}}=E_2^{\mathrm{t}}+T$ can be obtained by \cite[Lemma~1]{huleihel2016erasure} and the same argument as that used to justify    $E_1^{\mathrm{u}}=E_1^{\mathrm{t}}+T$ in Theorem~\ref{TH:E1}. \par 
 For constant composition random codes, by using the same argument in the end of the proof of Theorem \ref{TH:E1}, the result can be obtained. This concludes the proof of Theorem~\ref{TH:E2}.
 \end{IEEEproof}
\begin{remark}\label{rmk:two_max} {\em 
The maximization operations in~\eqref{eqn:E2_final_0} and~\eqref{eqn:E2_final} lead the somewhat unusual maximization in the error exponent $E_2^\rmt$ in~\eqref{eqn:E2} and hence~\eqref{def:ExpM2} in the theorem statement. We provide some intuition for it here. Recall the BSC example in Section~\ref{sec:iid_exp}. Note that the input distribution $P_{UX}$ is given and may be chosen in a sub-optimal manner so the regions in \eqref{capacityR1} and \eqref{capacityR2} are {\em not}   capacity regions.
\begin{itemize}
\item  If the  first inner minimization in \eqref{eqn:E2_final},   pertaining to  $f(\hatQ_{UXY})$ in \eqref{eqn:E2_final_0}, achieves the maximum,  this corresponds to terminal $\calY$  using the channel $W_\calY^n(y^n|X^n(m_1, m_2))$ to decode the true transmitted codewords $X^n(m_1,m_2)$ to find $m_2$. { On the other hand, if terminal $\calY$ decodes   $m_2$ successfully by using this option, roughly speaking, this corresponds to the event $\{F'_1\leq f_3e^{-nT}\}$ (see \eqref{eqn:E2_compact}) occurring. This case is analogous} to the rate constraint  $R_1 + R_2\le I(X;Y)$ in~\eqref{capacityR2}.

\item  If the second inner minimization in \eqref{eqn:E2_final}, pertaining to $s_0(\hat{Q}_{UY},R_1)$ in \eqref{eqn:E2_final_0}, achieves the maximum, this corresponds to $\calY$  using the induced channel $\Pr(y^n|\calC'_2(m_2))$ to decode { the sub-codebook $\mathcal{C}'_2(m_2)=\{X^n(m_1,m_2):m_1\in [M_1]\setminus\{1\}\}$. On the other hand, if terminal $\calY$ decodes the message $m_2$ successfully by using this option, roughly speaking,  this means that the event $\{ F'_1\leq F_2e^{-nT}\}$ (see \eqref{eqn:E2_compact}) occurs. This case is analogous}     to the rate constraint  $R_2\le I(U;Y) $ in \eqref{capacityR2}. 
\end{itemize} 
The {\em union} in \eqref{capacityR2} also corroborates the existence of the {\em maximum} in \eqref{eqn:E2_final}.}  
\end{remark}
 
\appendices
\section{Proof of proposition~\ref{prop:op1}}\label{pf:prop_op1}
\begin{IEEEproof}
Let $Q^*_1$ and $Q^*_2$ (two distributions of the form $Q=Q_{X|UY}\hat{Q}_{UY}$) be optimal solutions to the modified and original inner optimizations of $\Psi'_{\mathrm{b}1}$ and $\Psi_{\mathrm{b}1}$, respectively. Assume, to the contrary, that  $\beta(Q^*_2)>0$. {Moreover, note that $\beta(Q^*_1)<0$ by the assumption.} Due to the continuity of $\beta(Q)$ in $Q$, there exists a conditional probability distribution {$\barQ_{X|UY}$ such that $\bar{Q}=\alpha Q^*_1+(1-\alpha) Q^*_2$, where $\barQ=\barQ_{X|UY}\hat{Q}_{UY}$,} for some  $\alpha\in(0,1)$  that satisfies $\beta(\bar{Q})=0$, . As the first constraint in $\mathcal{L}_{21}$ is convex in $Q$, the solution $\bar{Q}$ is feasible (for $\Psi_{\mathrm{b}1}$). Note that the optimal value of objective function (in $\Psi_{\mathrm{b}1}$) is $\beta(Q^*_2)>0$ while $\beta(\bar{Q})=0$. This is a  contradiction. Hence, there exists an optimal solution to the original inner optimization problem $\Psi_{\mathrm{b}1}$ satisfying $\beta(Q_{X|UY}\hat{Q}_{UY})=0$. Moreover, this optimal solution of $\Psi_{\mathrm{b}1}$ (i.e., $(\hat{Q}^*,Q^*)$ with $\beta(Q^*_{X|UY}\hat{Q}^*_{UY})=0$) is also feasible for  $\Psi_{\mathrm{b}2}$. As a result, in this case, the optimal value of $\Psi_{\mathrm{b}}$ is equal to that for~$\Psi_{\mathrm{b}2}$ because $\Psi_{\mathrm{b}}=\min\{\Psi_{\mathrm{b}1},\Psi_{\mathrm{b}2}\}$. 
\end{IEEEproof}

\section{Proof of proposition~\ref{prop:op2}}\label{pf:prop_op2}
\begin{IEEEproof} 
Let $Q^*_1$, $Q^*_3$ and $Q^*_5$ (three distributions of the form $Q^*=Q^*_{XU|Y}\hat{Q}_{Y}$) be optimal solutions  to the modified and new  optimizations $\Phi'_{\mathrm{a}1}$, $\Phi'_{\mathrm{a}3}$ and $\Phi_{\mathrm{a}5}^*$, respectively. There are two cases for the solution $Q^*_5$, namely case (i) $\gamma(Q^*_5)\geq 0$ and case (ii) $\gamma(Q^*_5)\leq 0$.

In case (i), as the solution $Q^*_5$ is also optimal for the problem $\Phi'_{\mathrm{a}1}$, we only need to consider the solution $Q^*_3$ to the problem $\Phi'_{\mathrm{a}3}$. Note that the convex objective functions of $\Phi'_{\mathrm{a}3}$ and $\Phi_{\mathrm{a}5}^*$ are the same and the convex feasible set of $\Phi'_{\mathrm{a}3}$ is a subset of the convex feasible set of $\Phi_{\mathrm{a}5}^*$. Then the solution $Q^*_3$ must satisfy $\gamma(Q^*_3)=0$ by using a similar argument as that for $\Psi_{\mathrm{b}1}$ in the proof of Proposition~\ref{prop:op1}. Moreover, we may assume that this solution $Q^*_3$ is feasible for the original problem $\Phi_{\mathrm{a}3}^*$ (if not, similar to the discussion for $\Psi_{\mathrm{b}1}$, we do not need to consider this term $\Phi^*_{\mathrm{a}3}$ in~\eqref{opt1} due to $\Phi_{\mathrm{a}4}^*$). Hence, the optimal solution $Q^*_3$ to the problem $\Phi_{\mathrm{a}3}^*$ satisfies $\gamma(Q^*_3)=0$ and $\beta(Q^*_3)+0\geq 0$ is also feasible for the problem $\Phi_{\mathrm{a}1}^*$. Therefore, we can remove the term $\Phi^*_{\mathrm{a}3}$ in the inner minimization of~\eqref{opt1}.

For case (ii), using a similar argument as above, we can show that the solution $Q^*_1$ with $\gamma(Q^*_1)=0$ and $\beta(Q^*_1)\geq 0$ is feasible for the problem $\Phi_{\mathrm{a}3}^*$, therefore, we can remove this term $\Phi^*_{\mathrm{a}1}$ in the inner minimization of~\eqref{opt1}.

In summary, without loss of optimality, we can replace $\Phi_{\mathrm{a}1}^*$ and $\Phi_{\mathrm{a}3}^*$ with the new convex optimization problem $\Phi_{\mathrm{a}5}^*$. Moreover, the optimal value of $\Phi_{\mathrm{a}5}^*$ is {\em active} (see~\eqref{defact}) in the inner minimization problem in~\eqref{opt1} if 
\begin{align}
&\big\{\gamma(Q^*_5)\geq 0 \cap \beta(Q^*_5)\geq 0\big\}\bigcup \nn\\*
&\qquad\big\{\gamma(Q^*_5)\leq 0 \cap \gamma(Q^*_5)+\beta(Q^*_5)\geq 0\big\}.
\end{align}
This completes the proof of Proposition~\ref{prop:op2}.
\end{IEEEproof}

\section{Proof of Lemma~\ref{lem:remove_dependency}}\label{pf:lem_remove_dependency}
\begin{IEEEproof} 
{We are given  $(U^n(1),X^n(1,1),Y^n)=(u^n,x^n,y^n)$, $\calC'_U=c'_U$ and $T\geq 0$. Note also that $f_3$, defined in~\eqref{def:f3}, is constant/deterministic in this proof. In the following, we omit the dependence on the conditioning event $\{(U^n(1),X^n(1,1),Y^n)=(u^n,x^n,y^n),\calC'_U=c'_U\}$ for notational convenience.} Now we have 
\begin{align}
&\Pr\left\{{F_1+F_2}>\max\{f_3,F_4\}\cdot e^{-nT}\right\}\nonumber\\
&=\Pr\left\{{F_1+F_2}>F_4\cdot e^{-nT},F_4> f_3\right\}+\Pr\{F_4\leq f_3\}\nn\\*
&\quad\quad \times \Pr\left\{{F_1+F_2}>f_3\cdot e^{-nT}|F_4\leq f_3\right\}\\
&=\Pr\left\{{F_1+F_2}>F_4\cdot e^{-nT},F_4> f_3\right\}\nn\\*
&\qquad +\Pr\{F_4\leq f_3\}\cdot \Pr\left\{{F_1+F_2}>f_3\cdot e^{-nT}\right\}\label{newapp2}\\
&\doteq \Pr\left\{{F_1+F_2}>F_4\cdot e^{-nT},F_4> f_3\right\}+\Pr\{F_4\leq f_3\}\nn\\*
&\qquad \times \Pr\left\{\max\{F_1,F_2\}>f_3\cdot e^{-nT}\right\}\label{newapp3}\\
&\doteq\Pr\left\{{F_1+F_2}>F_4\cdot e^{-nT},F_4> f_3\right\}+\Pr\{F_4\leq f_3\}\nn\\*
&\qquad\times \max\left\{\Pr\{F_1>f_3\cdot e^{-nT}\},\Pr\{F_2>f_3\cdot e^{-nT}\}\right\}\label{newapp1} ,
\end{align}
where~\eqref{newapp2} is due to the fact that $(F_1,F_2)$ is independent of $F_4$ given $Y^n=y^n$ and $\calC'_U=c'_U$ (See the definitions of $F_1,F_2$ and $F_4$ in~\eqref{def:F1}--\eqref{def:F4}), {\eqref{newapp3} is due to the fact that $f_3\cdot e^{-nT}$ is exponentially small, and the interchange of $\max\{\cdot\}$ and $\Pr\{\cdot\}$ in \eqref{newapp1} is justified similarly as \cite[Eqn.~(37)]{somekh2011exact} and~\cite[Eqns.~(15)--(20)]{merhav2014exact}.} 

Recall the random codebook generation with superposition structure  as described in Section~\ref{sec:model}.  We may rewrite $F_1$ as
\begin{align}
F_1&= \sum_{m'_1\in \mathcal{M}_1\setminus\{1,2\}}\sum_{m'_2\in \mathcal{M}_2\setminus\{1\}} W_{\mathcal{Y}}^n(y^n|X^n(m'_1,m'_2))\nn\\*
&\qquad\qquad +\sum_{m'_2\in \mathcal{M}_2\setminus\{1\}}W_{\mathcal{Y}}^n(y^n|X^n(2,m'_2)). \label{eqn:F1_alt}
\end{align}
Note that  all  the $X^n(m_1, m_2)$ terms in $F_1$ and $F_4$  are generated in an i.i.d.\ manner. Hence the final term in~\eqref{eqn:F1_alt}, a non-negative random variable,  has the same distribution as $F_4$. 
Since  $W_{\mathcal{Y}}^n(y^n|\cdot)\geq 0$ and $ e^{-nT}\leq 1$, for any given $c'_U$, we obtain
\begin{align}\label{newlab3}
&\Pr\{F_1>f_3\cdot e^{-nT}\}
 \geq  \Pr\{F_1\! >\! f_3\}\! \geq\!  \Pr\{F_4\! > \! f_3\}.
\end{align}
The second inequality in~\eqref{newlab3} follows from the fact that if we have two random variables $A $ and $ A'$ which have the same distribution and $B$ is  a non-negative random variable, then clearly $\Pr\{A+B\ge c\}\ge\Pr\{A' \ge c\}$ for all $c\in\mathbb{R}$.
Now, we focus on the {sequence} $\eta_n :=\Pr\{F_4> f_3\}$. Assume that the limit of  $\eta_n $ exists (otherwise, we may pick a convergent subsequence and work with that subsequence in the following). There are two cases: case (i) $\lim_{n\to\infty} \eta_n =0$ and case (ii) $\lim_{n\to\infty} \eta_n >0$.\par
\begin{itemize}
\item For case (i), from~\eqref{newapp1}, we have:
\begin{align}
&\Pr\left\{{F_1+F_2}>\max\{f_3,F_4\}\cdot e^{-nT}\right\}\nonumber\\
&\doteq \Pr\left\{{F_1+F_2}>F_4\cdot e^{-nT},F_4> f_3\right\}\nn\\*
&\quad +\max\left\{\Pr\{F_1>f_3\cdot e^{-nT}\},\Pr\{F_2>f_3\cdot e^{-nT}\}\right\}\label{test1}\\
&\doteq \max\big\{\Pr\left\{{F_1+F_2}>F_4\cdot e^{-nT},F_4> f_3\right\},\nn\\*
&\qquad \Pr\{F_1>f_3\cdot e^{-nT}\},\Pr\{F_2>f_3\cdot e^{-nT}\}\big\}\\
&\doteq \max\left\{\Pr\{F_1>f_3\cdot e^{-nT}\},\Pr\{F_2>f_3\cdot e^{-nT}\}\right\},\label{newlab4}
\end{align}
where~\eqref{test1} is due to the fact that $\Pr\{F_4\leq f_3\}=1-\eta_n $ tends to 1, and the final step~\eqref{newlab4} is due to the fact that
\begin{align}
&\Pr\left\{{F_1+F_2}>F_4\cdot e^{-nT},F_4> f_3\right\}\nonumber\\
&\leq \Pr\{F_4>f_3\}\\
&\leq \Pr\{F_1>f_3\cdot e^{-nT}\}\label{pfappeq3} ,
\end{align}
and where~\eqref{pfappeq3} is due to~\eqref{newlab3}.
\item For case (ii), on the one hand, we have
\begin{align}
&\Pr\left\{{F_1+F_2}>F_4\cdot e^{-nT},F_4> f_3\right\}\nonumber\\
&\geq 1-\Pr\left\{{F_1+F_2}\leq F_4\cdot e^{-nT}\right\}-\Pr\{F_4\leq f_3\}\\
&\geq 1-\Pr\{F_1\leq F_4\cdot e^{-nT}\}-\Pr\{F_4\leq f_3\}\label{newapp4}\\
&\geq \Pr\{F_4> f_3\}-e^{-nR_1/4}\label{pfappeq1}\\
&\doteq 1\label{pfappeq4} ,
\end{align} 
where~\eqref{newapp4} is due to the fact that $F_2\geq 0$, \eqref{pfappeq1} is due to Lemma~\ref{lem:dependency} and~\eqref{pfappeq4} is due to the fact that~\eqref{pfappeq1} does not tend to $0$ (and is obviously bounded above by $1$) from the assumption that $\lim_{n\to\infty} \eta_n >0$. From~\eqref{newapp1}, we have 
\begin{align}\label{test3}
\Pr\left\{{F_1+F_2}>\max\{f_3,F_4\}\cdot e^{-nT}\right\}\doteq 1.
\end{align}
On the other hand, we have
\begin{align}
&\max\left\{\Pr\{F_1>f_3\cdot e^{-nT}\},\Pr\{F_2>f_3\cdot e^{-nT}\}\right\}\nonumber\\
&\geq \Pr\{F_1>f_3\cdot e^{-nT}\}\\
&\geq \Pr\{F_4> f_3\}\label{pfappeq2}\\
&\doteq 1\label{test2} ,
\end{align}
where~\eqref{pfappeq2} is due to~\eqref{newlab3}, and~\eqref{test2} is due to  the fact that~\eqref{pfappeq2} does not tend to $0$ from the assumption that $\lim_{n\to\infty} \eta_n >0$. Thus, for case (ii), combining~\eqref{test3} and~\eqref{test2}, we have 
\begin{align}
&\Pr\left\{{F_1+F_2}>\max\{f_3,F_4\}\cdot e^{-nT}\right\}\nn\\*
&\doteq  \max\left\{\Pr\{F_1>f_3\cdot e^{-nT}\},\Pr\{F_2>f_3\cdot e^{-nT}\}\right\}. \label{eqn:case2_conclu}
\end{align}
\end{itemize}
Since for both cases, we arrive at the same conclusions in~\eqref{newlab4} and~\eqref{eqn:case2_conclu}, this   completes the proof of Lemma~\ref{lem:remove_dependency}.
\end{IEEEproof}
\section{Proof of Lemma~\ref{lem:E3_vinish}}\label{pf:lem_E3_vinish}
\begin{IEEEproof}  The three parts of Lemma~\ref{lem:E3_vinish} are proved as follows:
\begin{enumerate}
\item Recall the definition of $\beta(Q_{UXY},R_1)$ (see \eqref{def:beta}). Let $Q'_{X|UY}=P_{X|U}$, we have
\begin{align}
&\beta(P_{X|U}\hatQ_{UY},R_1)\nn\\*
&=D(P_{X|U}\|P_{X|U}|\hatQ_U)+I_{P_{X|U}\hatQ_{UY}}(X;Y|U)-R_1\\
&=-R_1<0
\end{align}
Thus, there exists a conditional    distribution $Q'_{X|UY}=P_{X|U}$ belonging to 
$\{Q_{X|UY}:\beta(Q_{X|UY}\hatQ_{UY},R_1)\leq 0\}$.
\item As $E^*_3(\hatQ_{UY},t,R_1)$ is non-decreasing in $t$, we only need to show that  $E^*_3(\hatQ_{UY},t,R_1)=0$ when $t=s_0(\hatQ_{UY},R_1)$. From the conclusion above, we have $s_0(\hatQ_{UY},R_1)>-\infty$. Assume that the optimal solution corresponding to $s_0(\hatQ_{UY},R_1)$ is  $Q^*=Q^*_{X|UY}\hatQ_{UY}$. Now, we take $Q_{UXY}=Q^*$ for the constraint in $\mathcal{L}_4(s_0(\hatQ_{UY},R_1),R_1,R_1)$ in~\eqref{def:L4}, then we have
\begin{align}
&s_0+f(Q^*)-|R_1-\beta(Q^*,R_1)-R_1|_{+}\nonumber\\
&=[-f(Q^*)-\beta(Q^*,R_1)]+f(Q^*)-|-\beta(Q^*,R_1)|_{+}\\
&=[-f(Q*)-\beta(Q^*,R_1)]+f(Q^*)+\beta(Q^*,R_1)\label{eqn:s0_geq_t1}\\
&=0,
\end{align}
where~\eqref{eqn:s0_geq_t1} is because $Q^*$ satisfies the constraint $\beta(Q^*,R_1)\leq 0$ in~\eqref{eqn:s0_expression}. Therefore,  $Q^*\in \mathcal{L}_4(s_0(\hatQ_{UY},R_1),R_1,R_1)$ and $|\beta(Q^*,R_1)|_{+}=0$. Thus, $E^*_3(\hatQ_{UY},t,R_1)=0$ for $t\leq s_0(\hatQ_{UY},R_1)$.
\item Recall the definition $E_3(Q_{X|UY}\hatQ_{UY},t,R_1)$ and $E^*_3(\hatQ_{UY},t,R_1)$ (see \eqref{def:E3} and \eqref{def:E3_opt}). We only need to show that any conditional  probability  distribution $Q_{X|UY}$ such that $\beta(Q_{X|UY}\hatQ_{UY},R_1)\leq 0$ satisfies the condition  $Q_{X|UY}\notin \mathcal{L}_4(t,R_1,R_1)$. Assume, to the contrary, that there exist a conditional  probability distribution $\tilQ_{X|UY}$ such that $\beta(\tilQ,R_1)\leq 0$ and $\tilQ\in \mathcal{L}_4(t,R_1,R_1)$, where $\tilQ=\tilQ_{X|UY}\hatQ_{UY}$. Now, we have
\begin{align}
t&\leq -f(\tilQ)+|R_1-\beta(\tilQ,R_1)-R_1|_+\label{eqn:s0_leq_t1}\\
&=-f(\tilQ)-\beta(\tilQ,R_1)\label{eqn:s0_leq_t2}\\
&\leq \max_{\beta(\tilQ,R_1)\leq 0}[-f(\tilQ)-\beta(\tilQ,R_1)]\\
&=s_0(\hatQ_{UY},R_1)
\end{align}
where \eqref{eqn:s0_leq_t1} is because  $\tilQ\in \mathcal{L}_4(t,R_1,R_1)$ and \eqref{eqn:s0_leq_t2} is because  $\beta(\tilQ,R_1)\leq 0$. However, note that $t>s_0(\hatQ_{UY},R_1)$ as assumed in Lemma~\ref{lem:E3_vinish}. This is a contradiction. Hence, $E^*_3(\hatQ_{UY},t,R_1)$ is strictly positive for all $t>s_0(\hatQ_{UY},R_1)$.
\end{enumerate}
These  justifications complete the proof of Lemma~\ref{lem:E3_vinish}.\end{IEEEproof}

\subsubsection*{Acknowledgements}  The authors are indebted to the associate editor Prof.\ Neri Merhav and the two anonymous reviewers for extremely detailed comments that have helped to improve the clarity of the paper. The authors also thank Dr.\ Anshoo Tandon for discussions related to the examples in Section~\ref{sec:ex}.

\bibliographystyle{IEEEtran}
\bibliography{references}

\begin{IEEEbiographynophoto}{Daming Cao} received the B.Eng.\ degree in information
engineering from Southeast University, Nanjing, China, in 2013. He is currently working toward the Ph.D.\ degree from the School of Information Science
and Engineering, Southeast University. From Oct 2017 to Sep  2018, he was a visiting student in the Department of Electrical and Computer Engineering at the National University of Singapore. His research interests include information theory, network coding, and security.
\end{IEEEbiographynophoto}

\begin{IEEEbiographynophoto}{Vincent Y.\ F.\ Tan} (S'07-M'11-SM'15)  was born in Singapore in 1981. He is currently a Dean's Chair Associate Professor in the Department of Electrical and Computer Engineering  and the Department of Mathematics at the National University of Singapore (NUS). He received the B.A.\ and M.Eng.\ degrees in Electrical and Information Sciences from Cambridge University in 2005 and the Ph.D.\ degree in Electrical Engineering and Computer Science (EECS) from the Massachusetts Institute of Technology (MIT)  in 2011.  His research interests include information theory, machine learning, and statistical signal processing.

Dr.\ Tan received the MIT EECS Jin-Au Kong outstanding doctoral thesis prize in 2011, the NUS Young Investigator Award in 2014,  the Singapore National Research Foundation (NRF) Fellowship (Class of 2018) and the NUS Young Researcher Award in 2019. He is also an IEEE Information Theory Society Distinguished Lecturer for 2018/9. He has authored a research monograph on {\em ``Asymptotic Estimates in Information Theory with Non-Vanishing Error Probabilities''} in the Foundations and Trends in Communications and Information Theory Series (NOW Publishers). He is currently serving as an Associate Editor of the IEEE Transactions on Signal Processing.
\end{IEEEbiographynophoto}

\end{document}